\documentclass[12pt, preprint, longabstract]{aastex}
\usepackage{ccaption}

\shorttitle{OGLE-II bulge optical depth with bright sources}
\shortauthors{Sumi et al.}

\begin{document}

\title{Microlensing optical depth toward the Galactic Bulge using bright sources from OGLE-II}

\author{
T.~Sumi\altaffilmark{1}, 
P. R.~Wo\'zniak\altaffilmark{2}, 
A.~Udalski\altaffilmark{3}, 
M.~Szyma\'nski\altaffilmark{3}, 
M.~Kubiak\altaffilmark{3},
G.~Pietrzy\'nski\altaffilmark{3,4}, 
I.~Soszy\'nski\altaffilmark{3,4}, 
K.~\.Zebru\'n\altaffilmark{3}, 
O.~Szewczyk\altaffilmark{3},
\L.~Wyrzykowski\altaffilmark{3} \& 
B.~Paczy\'nski\altaffilmark{1}
}

\altaffiltext{1}{Princeton University Observatory, Princeton, NJ 08544-1001, USA,\\
e-mail: {\tt (sumi, bp)@astro.princeton.edu}}
\altaffiltext{2}{Los Alamos National Laboratory, MS-D466, Los Alamos, NM 87545,\\
e-mail: {\tt wozniak@lanl.gov}}
\altaffiltext{3}{Warsaw University Observatory, Al.~Ujazdowskie~4, 00-478~Warszawa, Poland,\\
e-mail: {\tt (udalski, msz, mk, pietrzyn, soszynsk, zebrun, szewczyk, wyrzykow)@astrouw.edu.pl}}
\altaffiltext{4}{Universidad de Concepci{\'o}n, Departamento de Fisica, Casilla 160--C, Concepci{\'o}n, Chile}

\begin{abstract}
%---------------

We present a measurement of the microlensing optical depth toward the Galactic Bulge
based on 4 years of the OGLE-II survey.  We consider only bright sources in the
extended Red Clump Giant (RCG) region of the Color-Magnitude Diagram, in 20 bulge fields 
covering $\sim5$ deg$^2$ between $0^\circ <l< 3^\circ$ and $-4^\circ <b< -2^\circ$.
Using a sample of 32 events we find
%$\tau=2.37_{-0.43}^{+0.53}\times 10^{-6}$ at $(l,b)=(1.^\circ16, -2.^\circ75)$.
% new 2.54826334436904e-06  -4.59692999125806e-07  5.66671277585328e-07
$\tau=2.55_{-0.46}^{+0.57}\times 10^{-6}$ at $(l,b)=(1.^\circ16, -2.^\circ75)$.
Taking into account the measured gradient along the Galactic latitude $b$,
%$\tau = [ (4.17\pm 2.20) + (0.73\pm 0.78)\times b]\times 10^{-6}$, this value is 
$\tau = [ (4.48\pm 2.37) + (0.78\pm 0.84)\times b]\times 10^{-6}$, this value is 
consistent with previous measurements using RCG sources and recent theoretical predictions. 
We determine the microlensing parameters and select events
using a model light curve that allows for flux blending.  Photometric quality delivered
by Difference Image Analysis (DIA) combined with the 1.3{\arcsec} median seeing of the OGLE-II
images are sufficient to constrain and reject the majority of strong blends.  We find that $\sim38$\%
of the OGLE-II events which appear to have RCG sources are actually due to much fainter stars blended
with a bright companion.  We show explicitly that model fits without blending result in similar
$\tau$ estimates through partial cancellation of contributions from higher detection efficiency, 
underestimated time-scales and larger number of selected events.  The near cancellation
of the optical depth bias and the fact that microlensing event selection based on models without
blending discriminates against blends have been utilized by previous analyses based on RCG sources.
The latter approach, however, leads to biased time-scale distributions and event rates.
Consequently, microlensing studies should carefully consider source confusion effects
even for bright stars.
\end{abstract}


\section{Introduction}
%=====================

Following the suggestion of \cite{pac91} and \cite{gri91} several teams have carried out microlensing surveys toward
the Galactic Bulge (GB). To date, well over $2\times10^3$ microlensing events in the GB have been detected
by those groups: OGLE (\citealt{uda94, uda00, woz01, uda03}), MOA (\citealt{bon01, sume03}),
MACHO (\citealt{alc97, alc00b}) and EROS (\citealt{afo03}). Thousands of detections are expected in the
upcoming years from MOA\footnotemark\footnotetext{\tt http://www.massey.ac.nz/\~{}iabond/alert/alert.html} and OGLE-III\footnotemark.
\footnotetext{\tt http://www.astrouw.edu.pl/\~{}ogle/ogle3/ews/ews.html}
It is now well understood that these observations are useful for studying the structure, kinematics and
dynamics of the Galaxy, and the stellar mass function, as the event rate and time-scale distributions are related
to the masses and velocities of lens objects.

The magnification of a microlensing event is described by (\citealt{pac86})

\begin{equation}
  \label{eq:amp-u}
  A(u)= \frac{u^2+2}{u\sqrt{u^2+4}},
\end{equation}

\noindent
where $u$ is the projected separation of the source and lens in units of the Einstein radius $R_{\rm E}$ which
is given by

\begin{equation}
  R_{\rm E}(M,x) = \sqrt{\frac{4GM}{c^2}D_{\rm s}x(1-x)},
  \label{eq:re}
\end{equation}

\noindent
where $M$ is the lens mass, $x=D_{\rm l}/D_{\rm s}$ is the normalized lens distance and $D_{\rm l}$ and
$D_{\rm s}$ are the observer-lens and the observer-source distances. The time variation of $u=u(t)$ is

\begin{equation}
  \label{eq:u}
  u(t)=\sqrt{u_{\rm min}^2 + \left( \frac{t-t_{0}}{t_{\rm E}} \right)^2},
\end{equation}

\noindent
where $u_{\rm min}$, $t_{0}$, $t_{\rm E}= R_{\rm E}/v_{\rm t}$ and $v_{\rm t}$ are, respectively, the minimum
impact parameter in units of $R_{\rm E}$, the time of maximum magnification, the event time-scale and the transverse
velocity of the lens relative to the line of sight toward the source star. From light curve alone, one can
determine the values of $u_{\rm min}$, $t_{0}$ and $t_{\rm E}$, but not the values of $M$, $x$ or $v_{\rm t}$.

Microlensing optical depth (the total cross section for microlensing) is directly related to the mass density
of compact objects along the line of sight (\citealt{pac96}). However, previous results have been controversial.
\cite{pac91} and \cite{gri91} first predicted the optical depth of $\tau\sim 5\times 10^{-7}$, assuming that all
events were associated with known disk stars. After the first several bulge events were reported by OGLE
(\citealt{uda94}), the high event rate prompted \cite{kir94} to evaluate the contribution  of bulge stars
in addition to the disk stars. They estimated $\tau\sim 8.5\times 10^{-7}$ and concluded that the value could
be about twice as large, if the bulge were elongated along the line of sight. Nevertheless, the first
measurements of the optical depth, $\tau\sim3.3 \times 10^{-6}$ by OGLE (\citealt{uda94}) and
$\tau\sim3.9^{+1.8}_{-1.2} \times 10^{-6}$ by MACHO (\citealt{alc97}), were well above the predictions.
Recent studies based on Difference Image Analysis (DIA)---less sensitive to the systematics of blending
in crowded fields---also found large optical depths: $\tau=3.23^{+0.52}_{-0.50} \times 10^{-6}$
at $(l,b)=(2.^{\circ}68, -3.^{\circ}35)$ from 99 events by MACHO (\citealt{alc00b}) and
$\tau=3.36^{+1.11}_{-0.81} \times 10^{-6}$ at $(l,b)=(3.^{\circ}0, -3.^{\circ}8)$ from 28 events by MOA
(\citealt{sume03}). The latter values were adjusted for the presence of the foreground disk stars, and the
uncorrected measurements are considerably lower, i.e. $\tau=2.43^{+0.39}_{-0.38} \times 10^{-6}$ and  
$\tau=2.59^{+0.84}_{-0.64} \times 10^{-6}$, respectively.

To explain high optical depths a number of authors have suggested the presence of a bar oriented along
our line of sight to the GB (\citealt{pac94}; \citealt{zha95}), and have adopted various values of the bar
orientation and mass (\citealt{han95}; \citealt{zha96}; \citealt{pea98}; \citealt{gyu99}). The resulting
values are in the range $\tau= 0.8 - 2.0 \times 10^{-6}$. \cite{bin00} have shown that high optical depth
measurements available at the time could not be easily reconciled with our general understanding of the
Galactic dynamics, and that the standard models of the Galaxy would need to be revised.

\cite{alc97} raised the possibility of a systematic bias in the optical depth measurement due to the
difficulties of measuring $t_{\rm E}$ associated with merging unresolved sources. When the actual source
base-line flux is unknown, $t_{\rm E}$ and $u_{\rm min}$ are degenerate in relatively low signal-to-noise
ratio (S/N) events (c.f. \citealt{woz97}; \citealt{han99}; \citealt{bon01}; \citealt{gou02}). \cite{pop01}
postulated that optical depth may be estimated without a bias due to blending by using only events with
bright source stars, such as Red Clump Giants (RCG), in which blending is assumed to be negligible.
Although the first measurements by \cite{alc97} gave a high value $\tau\sim3.9^{+1.8}_{-1.2} \times 10^{-6}$,  
the recent measurements based on events with RCG sources have returned lower
optical depths: $2.0 \pm 0.4 \times 10^{-6}$ at $(l,b)=(3.^{\circ}9, -3.^{\circ}8)$ from 50 events
by MACHO (\citealt{pop01}), $0.94 \pm 0.29 \times 10^{-6}$ at $(l,b)=(2.^{\circ}5, -4.^{\circ}0)$ from 16 
events by EROS (\citealt{afo03}) and $2.17^{+0.47}_{-0.38} \times 10^{-6}$ at
$(l,b)=(1.^{\circ}50, -2.^{\circ}68)$ from 42 events by MACHO (\citealt{pop04}).

In this paper we present a measurement of the microlensing optical depth in the direction of the GB based on
the full 4-year extent of the OGLE-II monitoring data. For analysis we select only high S/N events with
bright apparent source stars in the RCG region. In order to better understand the systematics of this complex
measurement, we proceed without assuming that blending is negligible. In the first part of the paper we present
the photometric data (\S\,\ref{sec:data}), selection of microlensing events (\S\,\ref{sec:select}),
computation of the detection efficiency (\S\,\ref{sec:eff}) and estimation of the optical depth
(\S\,\ref{sec:opt}). This is followed, in the second part (\S\,\ref{sec:disc} and \S\,\ref{sec:summary}),
by a description of various cross checks and general discussion. We primarily examine source confusion
and the effect of reintroducing the assumption of negligible blending in our bright event sample.

\section{Data}
%=============
\label{sec:data}

The data used in this analysis were collected between 1997 and 2000, during the second phase of the OGLE experiment.
All observations were made with the 1.3-m Warsaw telescope located at the Las Campanas Observatory, Chile.
The observatory is operated by the Carnegie Institution of Washington. The "first generation" camera has
a SITe 2048 $\times$ 2048 pixel CCD detector with pixel size of 24 $\mu$m resulting in 0.417 arcsec/pixel
scale. Images of the GB fields were taken in drift-scan mode at "medium" readout speed with the gain
7.1 $e^{-}$/ADU and readout noise of 6.3 $e^{-}$. A single 2048 $\times$ 8192 pixel frame covers an area
of 0.24 $\times$ 0.95 deg$^2$. Saturation level is about 55,000 ADU. Details of the instrumentation setup
can be found in \cite{uda97}.

In this paper we use $I$-band images for the 20 central OGLE-II fields in the GB.
Between 138 and 555 frames were available in each field. The remaining
29 fields are either harder to model (because of a prominent disk component), or cannot be treated as
co-located with the rest and averaged. The event number statistics in those fields is not sufficient
for an independent determination of $\tau$. A reliable map of the microlensing optical depth must await
the full analysis of the OGLE-III data. The centers of the analyzed fields are listed
in  Table~\ref{tbl:fld}. The time baseline of the survey is almost 4 years. There are gaps between the
observing seasons when GB cannot be observed from Earth, each about 3 months long. The median seeing
was $1.3''$. Astrometric and photometric scales are defined by multi-color maps of \cite{uda02},
comprising positions and $VI$-band photometry of $\sim 3\times10^7$ stars in all 49
GB fields. Photometric zero points are accurate to about 0.04 mag.

We adopt a hybrid approach to photometry with source detection and centroiding on reference images using DoPHOT
package (\citealt{sch93}), combined with the DIA photometry to lower the point-to-point
scatter (\citealt{ala98}; \citealt{ala00}; \citealt{woz00}). A high S/N reference image for each field
was taken to be a mean of 20 frames with the best overall seeing and background.
To obtain a difference frame we convolve the reference image with the PSF matching kernel and subtract
the result from a given survey frame, after interpolating to the same pixel grid. PSF photometry
on difference frames is performed only for objects detected on the reference image using fixed positions
found by DoPHOT.

\section{Microlensing event selection}
%=================================
\label{sec:select}

Criteria for selecting candidate microlensing events are summarized in Table~\ref{tbl:criteria}.

\subsection{Defining a population of bright Galactic Bulge sources}
\label{sec:level0}

For the optical depth analysis we select only those microlensing events that occurred in bright stars.
Red Clump Giants are sufficiently numerous for acceptable counting statistics, but not too numerous.
Their combined seeing disks still cover a small fraction of the survey area. Therefore, RCGs
should be less affected by source confusion with stars $\sim2$ mag deeper in the luminosity function (LF)
compared to fainter stars with higher number density. Note that by itself this does not guarantee that we can
entirely neglect blending in our samples. We discuss this issue in \S\,\ref{sec:level6} and \S\,\ref{sec:blending}.

\begin{figure}
\begin{center}
\includegraphics[angle=0,scale=0.8,keepaspectratio]{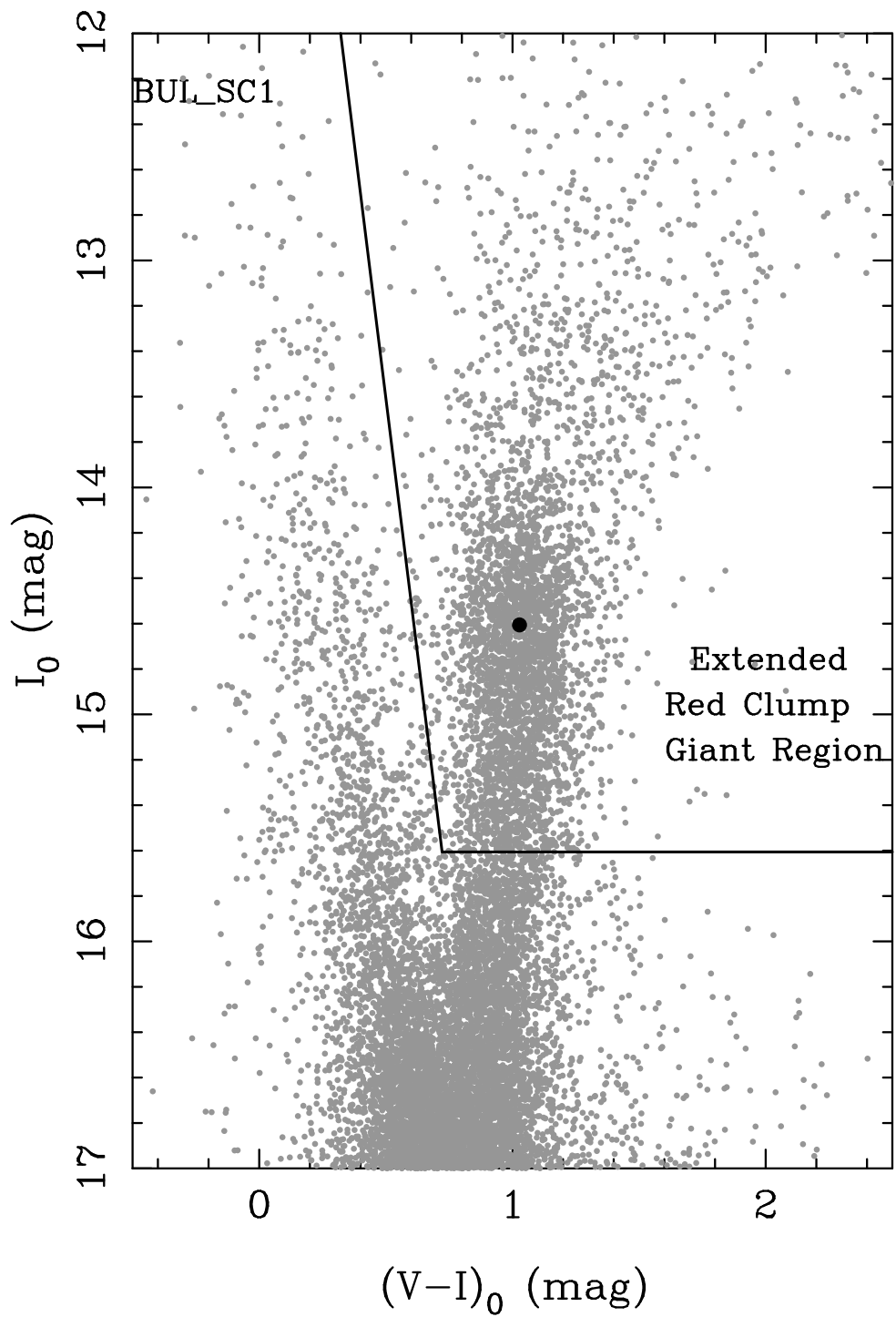}
\caption{
  \label{fig:cmd}
Color Magnitude Diagram of the BUL\_SC1 field. Our extended RCG region is defined as the upper right
portion of the figure with respect to the solid line. The filled circle represents the ``center''
of the RCG distribution from Sumi (2004).
}
\end{center}
\end{figure}

Only stars with more than 70 data points are considered.
In the Color Magnitude Diagram (CMD) we define the extended Red Clump Giant (RCG) region using a condition:
$I_0 < I_{\rm RC,th}$ and $ I_0 < 9 \times (V-I)_0 + I_{\rm RC}-5.5 $ (Fig.~\ref{fig:cmd}). Extinction
corrected magnitudes and colors, i.e. $I_0$ and $(V-I)_0$, are obtained for each star using
the OGLE-II reddening map of the GB (\citealt{sumEX04}). The magnitude threshold is set at $I_{\rm RC,th}=I_{\rm RC}+1$,
where $I_{\rm RC}$ is the mean magnitude of the RCG population in a given field. $I_{\rm RC}$ is estimated
according to the prescription in \cite{sumEX04} and accounts for the Galactic bar geometry. This procedure
is very effective in rejecting most of the disk main sequence. In contrast with previous analyses using RCG
sources (\citealt{afo03, pop04}), our extended RCG region includes not only RCGs, but also Red Giants and
Red Supper Giants. The total number of sources selected in all 20 fields was 1,084,267
(level 0 criteria in Table~\ref{tbl:criteria}).

\subsection{Searching for peaks consistent with microlensing}
\label{sec:level1and2}

Level 1 criteria detect generic flux brightening in otherwise quiet objects, followed
by microlensing model fitting at level 2 (Table~\ref{tbl:criteria}).
We define a "peak" as a group of consecutive data points in the light curve that includes
at least 5 data points with significance $\sigma \ge 1.6$. At least 3 of those points should have
$\sigma \ge 3$. Counting of consecutive data points allows for one failure, i.e. $\sigma < 1.6$
between data points.

The baseline flux and the significance of each photometric measurement with respect to that baseline
are established within a centered running window spanning $500$ d. We adopted:

\begin{equation}
  \label{eq:sigma_peak}
  \sigma_i = \frac{F_i - F_{\rm med,out}}{\sqrt{\sigma_{F,i}^2 + \sigma_{\rm out}^2}},
\end{equation}

\noindent
where $F_i$ and $\sigma_{F,i}$ are the flux and flux error of the $i$-th data point.
$F_{\rm med, out}$ and $\sigma_{\rm out}^2$ are the median and variance of all data
points outside the window.

There are further conditions on the number of detected peaks ($N_{\rm peak}$),
the maximum significance level reached by any point in any of those peaks
($\sigma_{\rm max}$), and the integrated significance of the peak where $\sigma_{\rm max}$ occurred
($\sum_{\rm peak,max} \sigma_i$). Full details are in Table~\ref{tbl:criteria}.
For lower S/N peaks ($\sigma_{\rm max} < 10$) we allow less noise in the baseline and limit the reduced
$\chi^2$: $\chi_{\rm out}^2/d.o.f. < 2.2$.

\begin{figure*}
\begin{center}
\includegraphics[angle=0,scale=0.8,keepaspectratio]{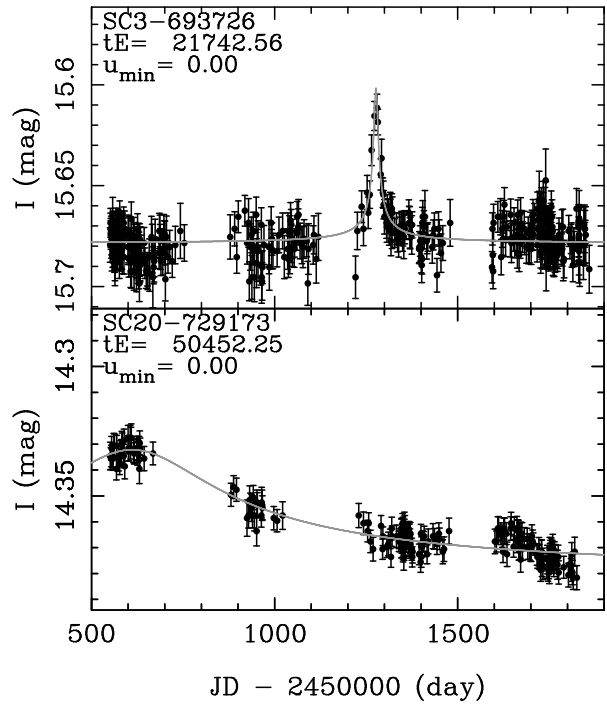}
\caption{Light curves of 2 sample events that were rejected by the condition $t_{\rm E}<400$ d.
The first object (top) has strongly degenerate and unreliable model parameters, and may
represent an extreme case of blending. The second object (bottom) is likely a low amplitude
variable star.
%The field/star ID, as well as the best fit parameters $t_{\rm E}$ and $u_{\rm min}$
\label{fig:lctE400} }
\end{center}
\end{figure*}

The non-linear model fitting is performed using the MINUIT package (\citealt{CERN98})
which provides the best fit parameters and corresponding 68\% confidence intervals
(for 1 parameter of interest). We fit a single lens microlensing model in the form:

\begin{equation} 
  \label{eq:deltaf} 
  F(t)= F_{\rm s} A(t) + F_{\rm b},
\end{equation}

\noindent
where $A(t)$ is given in Equations~\ref{eq:amp-u}--\ref{eq:u}, 
$F_{\rm s}$ and $F_{\rm b}$ are the baseline fluxes of the source star and the blended background
star, respectively. Table~\ref{tbl:criteria} lists all requirements on the quality of the best fit
model $\chi_{\rm ml}^2/d.o.f.$ and on the parameters. Long Period Variables and Novae in our data set
tend to have unusually large $t_{\rm E}$ or $u_{\rm min}>1$. None of the 135 candidates that remain
at this point are likely to be a variable star in one of those categories. 
In Fig. \ref{fig:lctE400} we show light curves of two objects which failed the $t_{\rm E}$ cut.

\subsection{Removing spurious events and low quality light curves}
\label{sec:level345}

PSF photometry on difference frames is performed assuming a constant background at zero counts.
The PSF matching kernel does not handle the extended wings of very bright stars due to lack
of signal there. This is why light curves of some stars follow the intrinsic or seeing induced
variability of their much brighter neighbors (\citealt{sumQSO04}). The effect is further exacerbated
by the presence of spurious objects around bright stars in DoPHOT output. Such artifacts around microlensed
stars and other variables (primarily Novae) have to be rejected from the list of microlensing events.
This is accomplished by cross-referencing the list of positions for our bright candidate events with an analogous
list constructed for events of any brightness (both after level 0--2 selection only).
In case there is a ``sibling'' event within 20 pixels that occurred at a similar time $t_0\pm6$ d,
the one with lower $\sigma_{\rm max}$ is rejected from its respective list.

Most of the 54 rejected light curves are
associated with a possible nova which appeared in BUL\_SC4 field around HJD=2450614.79 d. Stars
as far as 100 pixels around it were affected. Several rejected artifacts are associated with microlensing events.
At level 3 (Table~\ref{tbl:criteria}) our automated criteria select 81 candidate microlensing events.
All light curves in that group were also selected as microlensing events by visual inspection.
Up to this point, 41 candidates flagged as microlensing by eye were rejected because of low S/N or $u_{\rm min}>1$.
Note that strongly degenerate events can fail if they happen to have the best fit model with $u_{\rm min}>1$.

Out of the 81 remaining events, 69 have estimates of errors in all parameters from MINUIT. In practice, 
i.e. with any decent initial guess, the only reason for MINUIT error estimates not to converge is a 
very wide and shallow $\chi^2$ surface. Events with such degenerate models have unreliable best fit 
parameters and are not used in the optical depth measurement (level 4, Table~\ref{tbl:criteria}).
Furthermore, we required $\chi^2_{\rm ml}<3.5$ (level 5 selection in Table~\ref{tbl:criteria}). This rejects
three events with a low amplitude variability visible in the baseline and caused by the nearby bright variable stars
or by differential refraction. The requirement on the goodness of fit prevents such variability from distorting
the parameters of the fitted model.

\subsection{Exotic microlensing events}
\label{sec:level5}

Two binary lens events are rejected by the $\chi^2_{\rm ml}$ cut from the previous paragraph. Some binary microlensing
events show only weak deviations with respect to the basic case. The detectability of such events is sensitive
to the details of the temporal sampling and coverage. The deviations that were missed may still significantly distort
the results of the single lens model fitting. We cross-referenced our event sample with other published samples from
MACHO (\citealt{alc00a, tho04}), MOA (\citealt{sume03}), EROS (\citealt{afo03}, and the EROS alert system\footnotemark
\footnotetext{\tt http://dphs10.saclay.cea.fr/Spp/Experiences/EROS/alertes.html}), and OGLE (\citealt{jar02}).
We found two more binaries in addition to the ones that failed the $\chi^2_{\rm ml}$ condition:
SC21-45456 (97-BLG-1; \citealt{alc00a}), in which only the declining part of the light curve is covered by the OGLE data,
and SC20-69068 (sc20-1793; \citealt{jar02}), where the observed deviations are small. All four binary lens candidates
and their corresponding ID cross-references are listed in Table \ref{tbl:binary}. A complete list of cross-references
for our candidate events is available electronically on ApJ-web. 

In case of binary lens events it is difficult to find reliable lensing parameters like $t_{\rm E}$ and $I_s$ without very
good coverage. Unfortunately, that is the case for most of our candidates. Furthermore, it is a complicated task to estimate
the detection efficiency for binary microlensing. The fraction of binary lens events is around 3--8\% (\citealt{alc00a,jar02,jar04}).
In \S\,\ref{sec:opt} this fraction is used to account for the optical depth contribution from binary lens events.
This avoids estimating $\tau$ due to individual binary lenses, which is subject to large uncertainties in our sample.

One of the events, namely SC33-553617 (sc33\_4505 in \citealt{smi02}),
is a strong parallax event, and is included in the optical depth measurement in \S\,\ref{sec:opt}.
None of the remaining candidate parallax events listed in \cite{smi02} are part of our sample, including
possible parallax events with weak signatures.

\subsection{Rejecting events with faint, strongly blended sources}
\label{sec:level6}

\begin{figure}
\begin{center}
\includegraphics[angle=-90,scale=0.6,keepaspectratio]{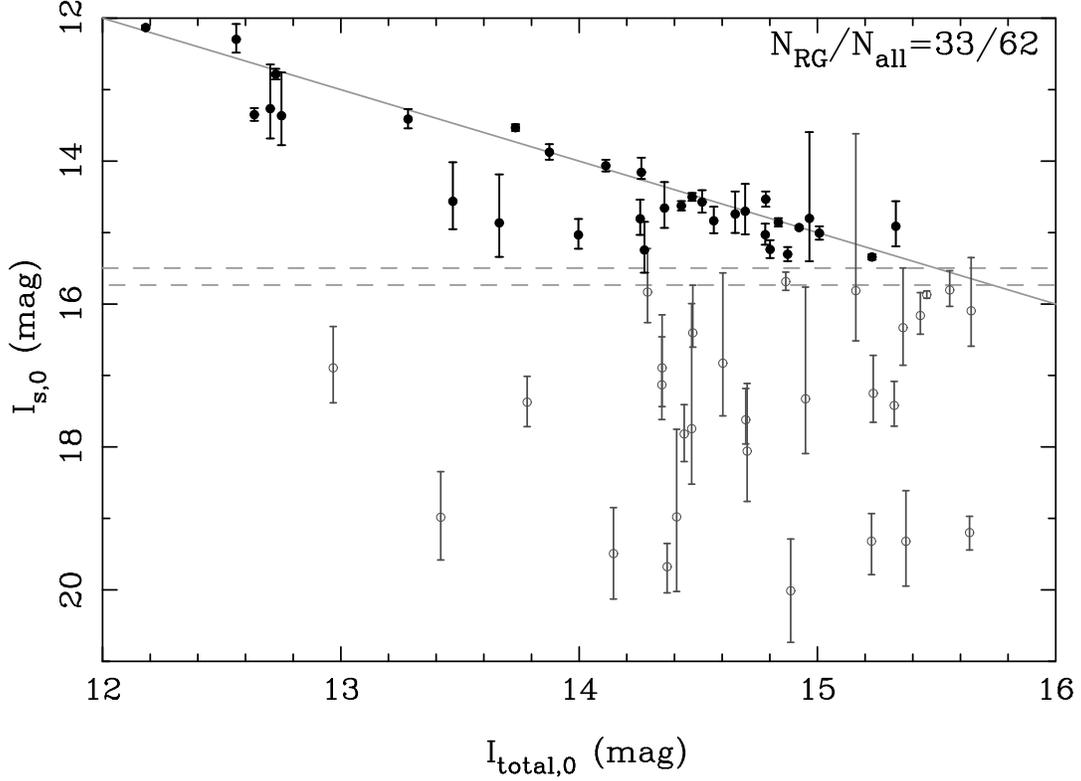}
\caption{Best fit source magnitude $I_{\rm s,0}$ versus the total
baseline magnitude $I_{\rm total,0}$ for 62 OGLE-II microlensing events.
Magnitudes are corrected for extinction.
Filled and open circles represent, respectively, sources above and below the threshold
at $I_{\rm RC,th}$ (c.f. the final cut in Table~\ref{tbl:criteria}).
Solid line indicates $I_{\rm s,0}=I_{\rm total,0}$, a blending free case.
Dashed lines indicate the range of field dependent threshold $I_{\rm RC,th}$.
\label{fig:IsItotal} }
\end{center}
\end{figure}

In this section we look for evidence of blending using only information from light curve fitting. Model degeneracy
is known to be a severe problem (\citealt{woz97}; \citealt{han99}; \citealt{bon01}; \citealt{gou02}). It was therefore
of critical importance, that all 62 events in our level 5 sample have
exceptionally high S/N ratio and very good coverage. This allowed us to make use of the fitted lensed flux
fraction $f_{\rm S}=F_{\rm s}/(F_{\rm s}+F_{\rm b})$ and its error. Fig.~\ref{fig:IsItotal} shows the resulting relation between
the apparent magnitude of the total light ($I_{\rm total,0}$) and the best fit source magnitude ($I_{\rm s,0}$), both corrected
for extinction. Considering a positive error range $\sigma_{I_{\rm s+}}$ we can construct an upper limit for the source magnitude
as $I_{\rm s,0} + \sigma_{I_{\rm s+}}$.
This leads to an unexpected conclusion that, with high probability, in 29 out of 62 events the source may actually be below
$I_{\rm RC, th}$ and does not satisfy level 0 conditions. In other words, the actual source in such an event
merged with a much brighter star to make a single PSF peak detected by the DoPHOT code. For such events the detection efficiency
cannot be estimated assuming one possible lensed star per database light curve. We remove those 29 events
(level 6, Table~\ref{tbl:criteria}) and for the optical depth estimate we use only the remaining 33 high quality events
with low probability of having $I_{\rm s,0} > I_{\rm RC, th}$. There are 32 unique events because SC31-111306 and SC30-717699
are the same event found in the overlap region between two fields. Here we limit the discussion to light curves and selecting
the final event sample. Our unexpected result prompted a much more rigorous analysis of blending and a detailed discussion
is deferred to \S\,\ref{sec:blending}.

We note a relatively large number of events detected in field BUL\_SC30 compared to the neighboring fields SC22 and SC23.
\cite{pop04} detected a similar excess at a location about 1 deg away. It remains to be established whether these
fluctuations imply a statistically significant clustering of microlensing events.

Light curves of 33 microlensing events in the final optical depth sample are shown in Fig.~\ref{fig:lc}, and their
best fit parameters are listed in Table~\ref{tbl:candlist6}. In case of SC33-553617 we also fitted a parallax model,
in addition to a standard single lens model. The best fit parallax model is shown as dotted line in Fig.~\ref{fig:lc}
and its parameters are given in Table~\ref{tbl:candlist6}. Using the notation from \cite{sos01} we found $\psi= 3.17\pm0.12$ rad
and $\tilde{r}_{\rm E}=6.16\pm0.39$ AU for the heliocentric orientation of the lens trajectory and the Einstein ring radius
at the observer plane, respectively. These values differ by about 2.7 $\sigma$ from results in \cite{smi02}. The discrepancy
is insignificant and fully accounted for by slight differences in the selection of good science frames and the details of
photometric calibrations.

\begin{figure}
\begin{center}
\includegraphics[angle=0,scale=0.8,keepaspectratio]{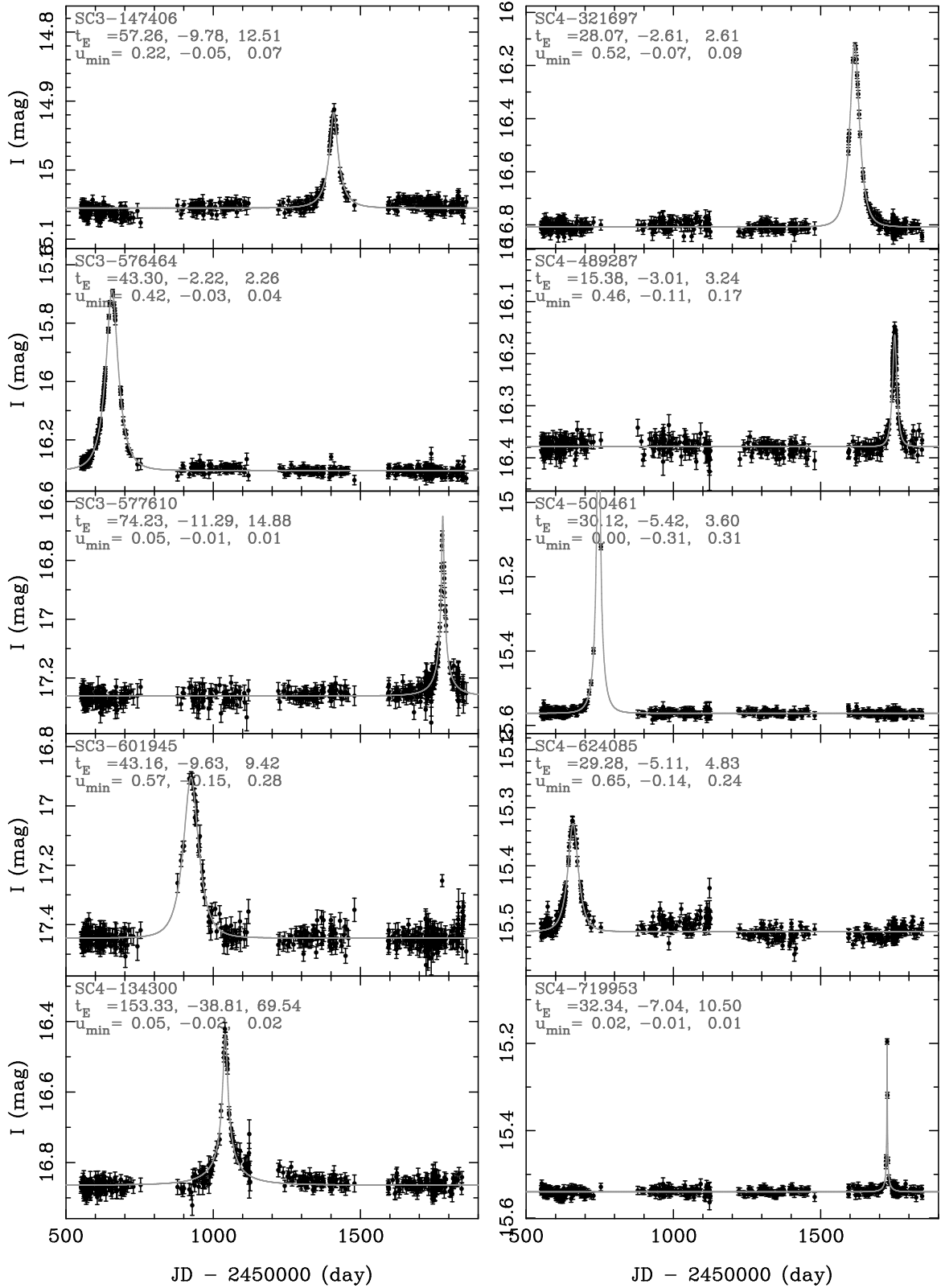}
\caption{Sample light curves of 10 level 5 events rejected for blending by level 6 criteria.
  \label{fig:lclevel5} }
\end{center}
\end{figure}

In Table~\ref{tbl:candlist5} we list level 5 events that were rejected for blending at level 6. 
Examples of light curves from the group showing evidence for blending are plotted in Fig.~\ref{fig:lclevel5}. 
For better readability Tables~\ref{tbl:candlist6} and \ref{tbl:candlist5} only contain symmetric (parabolic) 
68\% confidence level intervals. The analysis uses proper asymmetric confidence intervals for 1 parameter of interest.
Complete information on model fits and their uncertainties is available in the electronic form from ApJ web.
Supplemental files include a complete list of cross-references of our microlensing events with other published work,
as well as other parameters and event samples that could not fit comfortably in this paper.

\section{Detection efficiency}
%==================================
\label{sec:eff}

We estimate this crucial ingredient of the microlensing optical depth measurement in a two-step process. In the first part
we construct a synthetic reference image for each field to simulate blending of stellar flux and the effects of source confusion
on object catalogs in OGLE-II (see \S\,\ref{sec:blending} for a detailed discussion of blending). The second part deals with the
ability to detect microlensing events in database light curves known as sampling efficiency.
This reflects the way our actual data set was processed. Stars were detected and their positions were measured on reference images
using the DoPHOT code. In order to reduce photometric noise the DIA software was later used to measure fluxes. The transformation of the
difference flux to standard magnitudes also involves DoPHOT photometry, which determines the baseline flux.

\subsection{Blending distributions from synthetic images}
\label{sec:im_eff}

Synthetic stars were drawn from a luminosity function (LF) derived separately for each field. We made composite LFs with
the bright end based on the photometric maps of \cite{uda02}. The faint end of the LF, in all fields, was taken to be the same
as the HST LF in the Baade's Window (\citealt{hol98}), except for adjustments due to variable field extinction and Galactic bar
geometry (compare \citealt{sumEX04}). Both pieces have good completeness near $I_0=16$ mag, where they were joined and normalized.
We constructed synthetic reference images starting from a flat Poisson background. Those images were further populated
with stars at random positions down to the extinction free magnitude $I_0\sim21.0$ mag, matching the observed number
density of bright stars in the field. The PSF flux of each star was reduced according
to the extinction map in \cite{sumEX04} interpolated at the appropriate position from a $\sim30''$ resolution.
Local PSF shapes were also determined on a star by star basis using spatial PSF models from \cite{woz00}. 
Poisson noise was added in the amount that approximated the statistics of real reference frames.
Those simulated frames were processed with DoPHOT and the output
source catalog was constructed using procedures identical to those in the preparation of the real database. For each simulated
star in the input list we found the closest star in the output DoPHOT catalog. Obviously, the number of detected stars
is much lower than the number of simulated stars due to blending, but also because at some low flux level the GB stars
are so numerous that they form a smooth background.

\begin{figure}
\begin{center}
\includegraphics[angle=-90,scale=0.7,keepaspectratio]{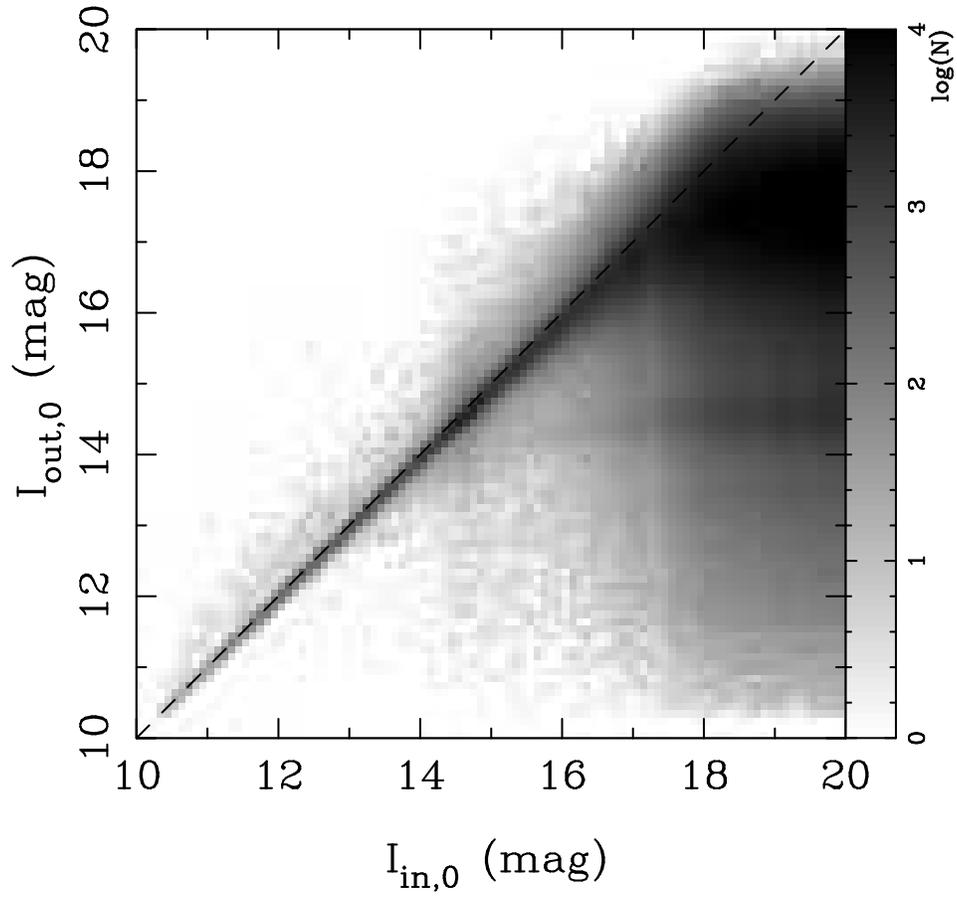}
\caption{Number density of the relation between the input ($I_{\rm in,0}$)
and the output ($I_{\rm out,0}$) extinction corrected magnitudes
simulated for BUL\_SC3 field. Source confusion intrinsic to the simulation
is visible in a small fraction of detections located just above the diagonal line.
\label{fig:IoutIin} }
\end{center}
\end{figure}

\begin{figure}
\begin{center}
\includegraphics[angle=-90,scale=0.6,keepaspectratio]{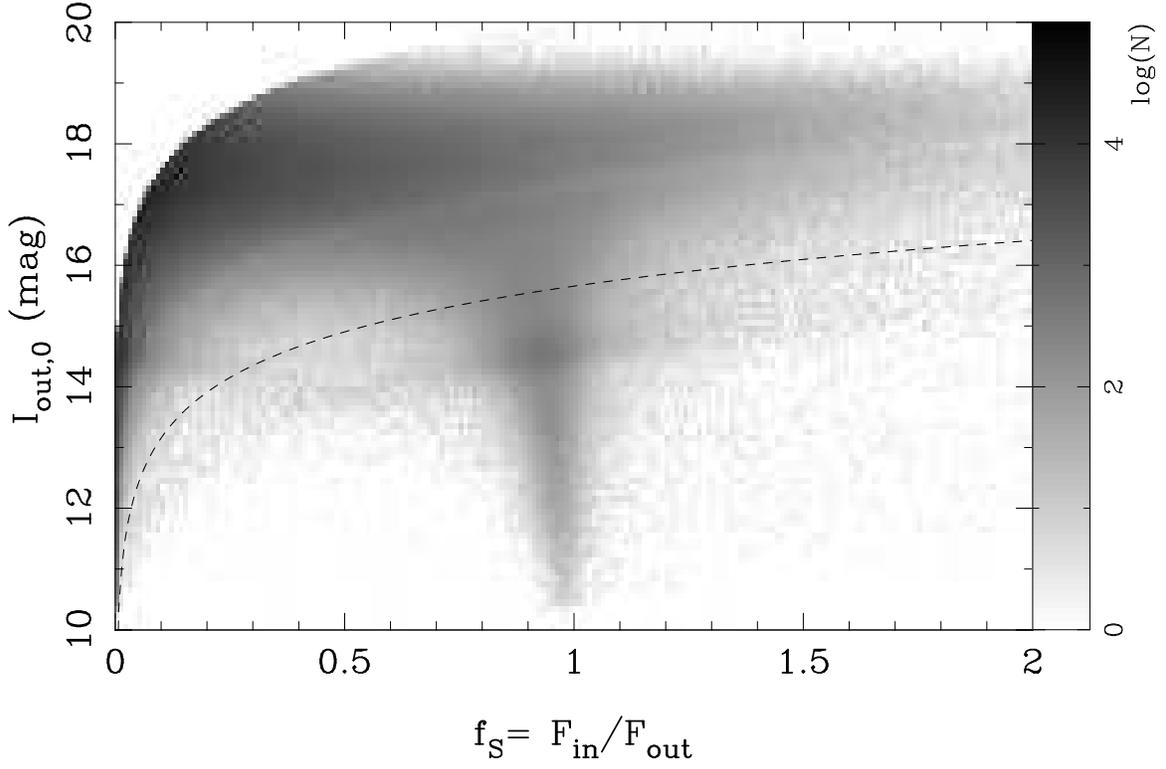}
\caption{Number density of the blending parameter $f_{\rm S}$ as a function
of the extinction corrected output magnitude $I_{\rm out,0}$
simulated for BUL\_SC3 field. The luminosity function is truncated for objects
fainter than $I_{\rm in,0}=20$ mag. The area below the dashed line represents true sources
brighter than the threshold: $I_{\rm in,0}<I_{\rm RC,th}=15.66$ mag.
  \label{fig:IfS} }
\end{center}
\end{figure}

\begin{figure}
\begin{center}
\includegraphics[angle=-90,scale=0.6,keepaspectratio]{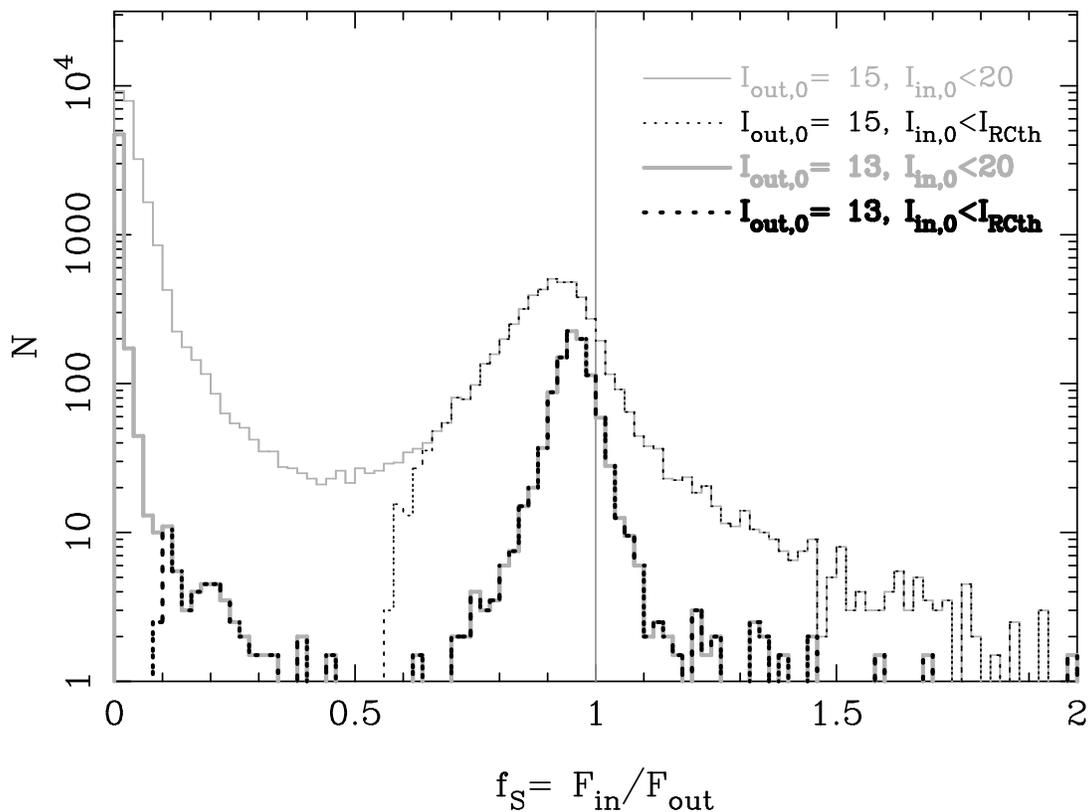}
\caption{Histograms of the blending parameter $f_{\rm S}$ in two narrow ranges of the extinction
corrected output magnitude $I_{\rm out,0}$, as simulated for BUL\_SC3 field:
$15-15.2$ mag (thin lines) and $13-13.2$ mag (thick lines).
Solid and dashed lines are for $I_{\rm in,0}<20$ mag and $I_{\rm in,0}<I_{\rm RC,th}=15.66$ mag, respectively.
  \label{fig:histfS} }
\end{center}
\end{figure}

In Fig.~\ref{fig:IoutIin} we show the relation between the extinction corrected $I$ magnitude of the input star ($I_{\rm in,0}$)
and the closest output star ($I_{\rm out,0}$). One can see that a large majority of bright stars is recovered correctly,
however great numbers of faint simulated stars ``come out'' as random bright stars. A faint star has a very high
probability of blending with a bright star to form an even brighter catalog source. This is best studied by Monte-Carlo
simulations because the ability to separate close stars in the photometric code depends in a complicated way at least on
the flux, the flux ratio, the separation and the PSF shape. Each star from the input LF has equal probability
of microlensing, so Fig.~\ref{fig:IoutIin} can be transformed into a relation between the blending parameter $f_{\rm S}$
and $I_{\rm out,0}$ as shown in Fig.~\ref{fig:IfS}. This is simply a prior probability that the lensed flux is a fraction
$f_{\rm S}$ of the total flux,
given a random event with the observed magnitude of the composite $I_{\rm out,0}$. Histograms of $f_{\rm S}$
were obtained by binning the map in Fig.~\ref{fig:IfS} into 0.2 mag intervals of $I_{\rm out,0}$, and were used in the next
step to generate random events in the efficiency simulation on database light curves. Fig.~\ref{fig:histfS} presents some examples.

It must be stressed that our blending distributions are subject to some (small) uncertainties intrinsic to the simulation.
IDs of merging objects need to be traced between the simulated input and output catalogs, which does not take place during the
construction of the photometric database (\S\,\ref{sec:data}). The exact allocation of flux to detected objects
is affected by the finite accuracy of positions and the photon noise interacting with the deblending code.

\subsection{Light curve sampling efficiency}
\label{sec:cat_eff}

In order to simulate real sampling and photometric accuracy we ``inject'' additional flux and photon noise into actual light
curves, randomly selected from the group searched for microlensing (\S\,\ref{sec:select}). Simulated events have random input
parameters drawn from flat distributions: $0<u_{\rm min}<1$, $2450530<t_0<2451860$ JD and
$\log({\rm 1\,d})<\log(t_{\rm E})<\log({\rm 500\,d})$. The blending parameter $f_{\rm S}$ is drawn from one of the histograms
made in the previous section according to the baseline magnitude (Fig.~\ref{fig:histfS}). The shape of this histogram varies
depending on how deep we integrate in $I_{\rm in,0}$. The efficiency for faint stars is low and very time consuming to compute, so
we looked at possible improvements. A preliminary simulation was conducted for BUL\_SC3 (one of the densest OGLE-II fields)
by generating 20,000 events down to $I_{\rm in,0} < 20$. We strictly repeated the event selection process from \S\,\ref{sec:select}.
Figs.~\ref{fig:fS} and \ref{fig:Isc} compare the most important simulation results to observed distributions. In \S\,\ref{sec:blending}
we will use this information to better understand blending effects.

\begin{figure}
\begin{center}
\includegraphics[angle=-90,scale=0.6,keepaspectratio]{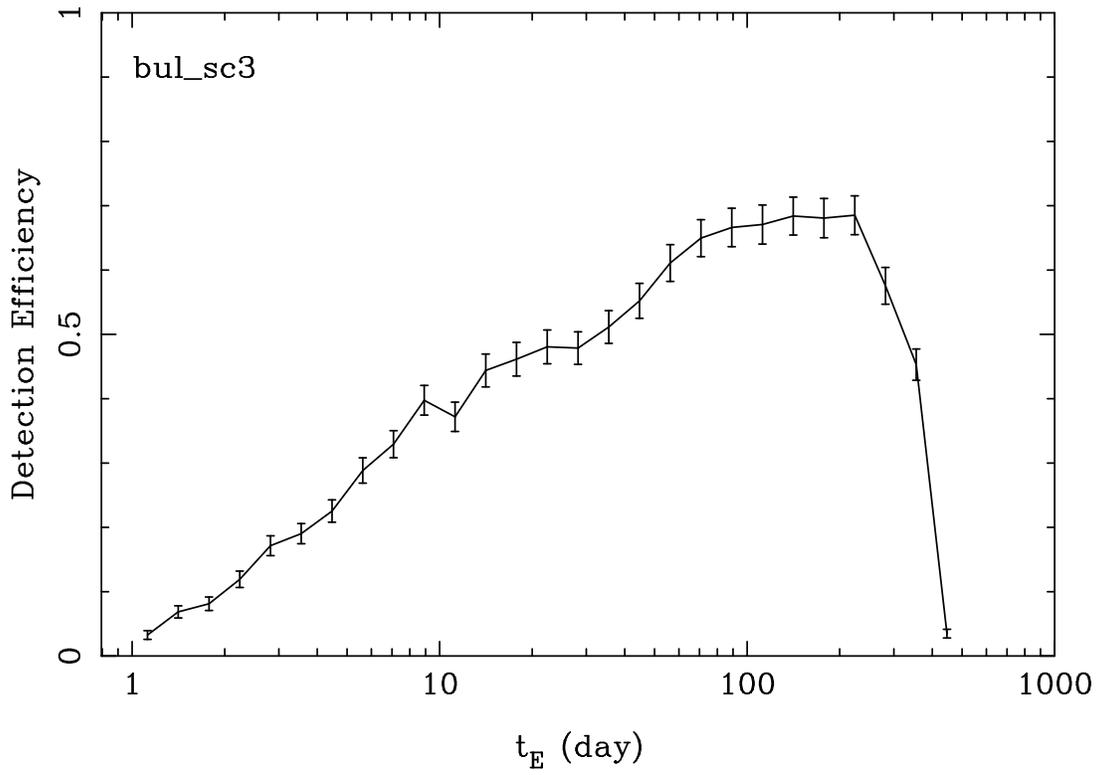}
\caption{OGLE-II microlensing detection efficiency as a function of $t_{\rm E}$ in BUL\_SC3 field.
  \label{fig:eff} }
\end{center}
\end{figure}

From the data shown in the top-right panel of Fig.~\ref{fig:Isc} we find that only 1\% of events in
the simulated optical depth sample have input source stars fainter than our brightness threshold $I_{\rm RC,th}$.
We verified that this fraction remains very low for other distributions of $t_{\rm E}$. Then we can safely assume
$I_{\rm in,0}<I_{\rm RC,th}$ when constructing the histograms of $f_{\rm S}$ for subsequent simulations, i.e. we use stars below
the dashed line in Fig.~\ref{fig:IfS}. It means that our simulated source LF is the one actually measured by the OGLE-II experiment,
making our calculations very insensitive to uncertainties in the faint part of the LF. (Sufficiently deep HST data are currently
available only for Baade's window.) The actual detection efficiency in each field is computed from 20,000 simulated events with
blending down to $I_{\rm RC,th}$ and integrated over all parameters except $t_{\rm E}$. The results for BUL\_SC3 field
are plotted in Fig.~\ref{fig:eff}. They are very similar for other fields,
which are available electronically on ApJ-web. 
These efficiencies are reduced by about
10\% when compared to the case without blending, because the main peak in $f_{\rm S}$ is near 0.9 for most stars in our RCG region.

\section{Microlensing optical depth}
%==================================
\label{sec:opt}

The optical depth $\tau$ can be defined as the probability that a random star is microlensed with the impact parameter
$u_{\rm min} \le 1$ at any given time. It can be estimated as

\begin{equation}
  \label{eq:opt}
  \tau = \frac{\pi}{2N_{\rm s}T_{\rm o}} \sum_i \frac{t_{{\rm E},i}}{\varepsilon (t_{{\rm E},i})},
\end{equation}

\noindent
where $N_{\rm s}$ is the total number of source stars monitored for microlensing, $T_{\rm o}$ is the duration
of the survey in days, $t_{{\rm E},i}$ is the event time-scale for the $i$-th event, and $\varepsilon (t_{{\rm E},i})$
is the detection efficiency at that time-scale. In our case $N_{\rm s}=1,084,267$ and $T_{\rm o}=1330$ d.
% 1860.0-530.0 =1330

Blending simulations in \S\,\ref{sec:im_eff} show that stars below our magnitude threshold have practically no chance
to contribute detectable events. This allows us to simplify and speed up the calculation by only simulating stars with
$I_{\rm s, 0} < I_{\rm RC, th}$, but also has a side effect of giving about average efficiency to a small percentage
of simulated stars which crossed the threshold and have practically no efficiency in the actual survey. Equivalently,
with the efficiency computation from \S\,\ref{sec:cat_eff}, the number of sources that can be lensed is slightly overestimated
due to blending. In principle, the effect can be corrected by lowering $N_{\rm s}$ in Equation~\ref{eq:opt} according to the
ratio between the input and output LFs integrated above $I_{\rm RC, th}$. Depending on the field, the resulting optical depth
would increase by 7--12 \%, or $\sim 9\%$ on average. The size of the correction is uncertain by a factor $\sim1.5$ due to
several competing
biases at a few percent level. To name a few, they include: limited accuracy and source confusion intrinsic to our image
simulations (\S\,\ref{sec:im_eff}), occasional DoPHOT artifacts in the database, and second order effects like not counting
in $N_{\rm s}$ a tiny fraction of ``balanced'' blends above the magnitude threshold (those with $f_{\rm S}\sim 0.5$ and near
the bright end of the LF; see \S\,\ref{sec:centroid_blends}).  Over-counting of RCG sources is partially canceled
by about 2\% due to artificial events having $I_{\rm s, 0} > I_{\rm RC, th}$, an effect caused by using 0.2 mag bins
to store the $f_{\rm S}$ distribution (see Fig. \ref{fig:histfS}).  Those stars cannot produce detectable events in our
simulation.  In the end, we rescaled the optical depth and the error by $9-2=7$\%.

According to the level 5 criteria (Table~\ref{tbl:criteria}), all four binary lens events were removed from the optical depth
sample. The model parameters of those events (including $t_{\rm E}$) are poorly constrained, and their RCG membership
is uncertain. The fraction of binary lens events among all microlensing events has been estimated at 8\% (\citealt{jar02}),
6\% (\citealt{alc00a}), and 3\% (\citealt{jar04}). In our RCG sample, we found evidence for a binary lens fraction $4/62=0.06$,
in good agreement with previous work.
We use this value to correct our optical depth measurement for binary lens events excluded from the sample.
Assuming that the lens system consists
of two stars having the same typical time scale, the optical depth contribution of a binary lens event is $2^{1/2}$ times that of
a single lens event. The ratio between the number of events with binary lenses and single lenses is $0.06/(1-0.06)$. 
It follows that the optical depth values and their errors have to be rescaled by a factor 1.09.

Individual optical depth estimates for each field are listed in Table~\ref{tbl:fld}.
The errors were estimated using the formula of \cite{han95}. We also estimated the average
optical depth in all 20 fields combined, and found 
%$\tau = 2.37_{-0.43}^{+0.53} \times 10^{-6}$ at $(l,b)=(1.\arcdeg 16, -2.\arcdeg 75)$.
$\tau=2.55_{-0.46}^{+0.57}\times 10^{-6}$ at $(l,b)=(1.^\circ16, -2.^\circ75)$.
The effective line of sight was computed by weighting field centers with the number of source stars in each field.
Here the error was estimated using a Monte-Carlo method from \cite{alc97}, which is
consistent with a value $\pm 0.47$ from the formula of \cite{han95}.

\begin{figure}
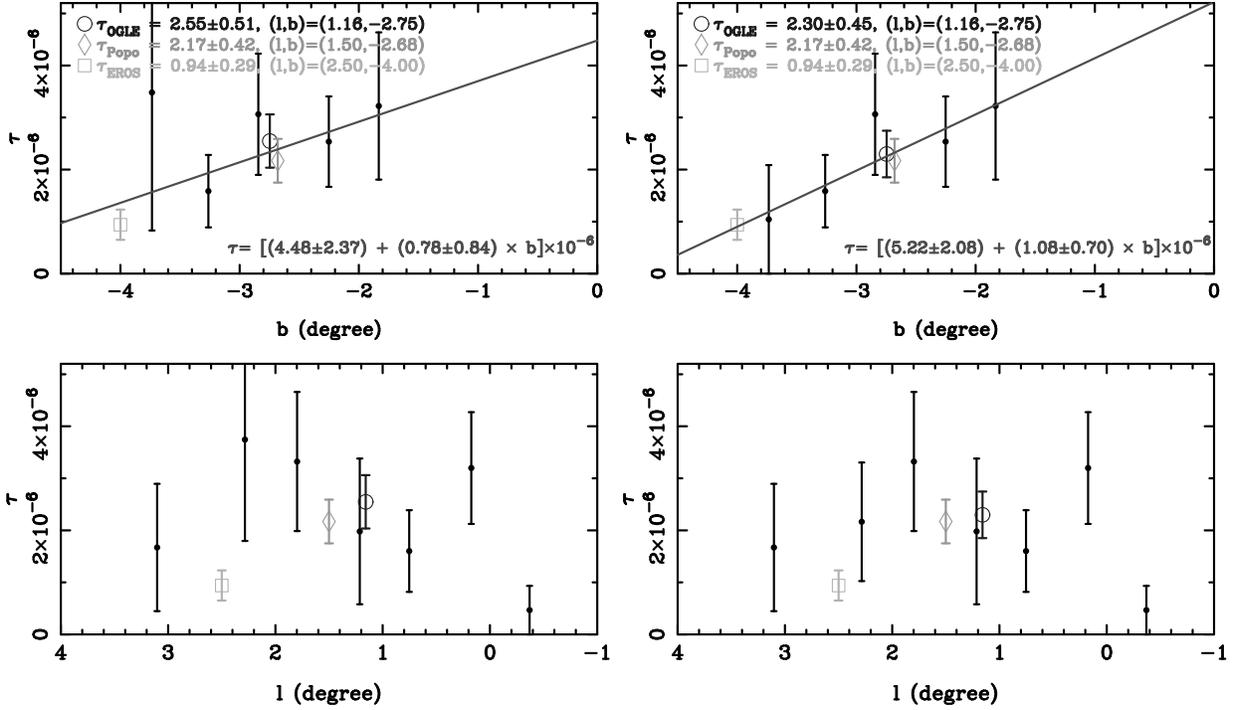

\begin{center}
\includegraphics[angle=0,scale=0.45,keepaspectratio]{f12a.eps}
\includegraphics[angle=0,scale=0.45,keepaspectratio]{f12b.eps}
\caption{Microlensing optical depth $\tau$ as a function of the Galactic $l$ and $b$ (filled circles). The results
based on the full level 6 sample from OGLE-II (left) are compared to the ones obtained after the parallax event
has been removed (right). The latter case is a better representation of the bulge structure since the parallax
event was likely caused by a disk lens. Also shown is the average $\tau$ in all 20 fields weighted by the
stellar number density (open circle). Solid line is the best fit linear gradient along $b$.
There is no indication of gradient along $l$ in this data.
Open diamond and open square represent results from Popowski et al. (2004) and Afonso (2003).
\label{fig:opt}
}
\end{center}
\end{figure}

The left panels of Fig.~\ref{fig:opt} show the microlensing optical depth from OGLE-II bulge data as a 
function of the Galactic coordinates $b$ and $l$ (filled circles). Fields were binned in ranges 
$\Delta b, \Delta l = 0.5^{\circ}$ with $l$ and $b$ averaged in each bin.
The errors are estimated according to \cite{han95}. Although there is no correlation in $l$, we can see 
a trend in the optical depth along $b$. A simple linear fit gives 
%$\tau = [ (4.17\pm 2.20) + (0.73\pm 0.78)\times b]\times 10^{-6}$.
$\tau = [ (4.48\pm 2.37) + (0.78\pm 0.84)\times b]\times 10^{-6}$.
In Fig.~\ref{fig:opt} we also compare the average $\tau$ from this work (open circle) with values published by \cite{pop04}
(open diamond) and by \cite{afo03} (open square). All three measurements are roughly consistent.
They fall well below the results in \cite{alc00b} and \cite{sume03} based on all survey stars regardless of
brightness ($\tau \sim3.3 \times 10^{-6}$ at $b\sim-3.^{\circ}5$). However, the latter values include a dubious upward
correction by about 20\% for the presence of disk stars that rarely produce events, and the uncorrected estimates
($\tau=2.43^{+0.39}_{-0.38} \times 10^{-6}$ and $2.59^{+0.84}_{-0.64} \times 10^{-6}$, respectively)
are only $\sim1.5\sigma$ higher than our present result. When comparing estimates made by various experiments,
we adjust for the measured gradient of $\tau$. Our new optical depth value is also consistent with predictions
based on the revised COBE bar model by \cite{han95}, which has the mass 
$M_{\rm bulge}=1.62\times 10^{10} M_\odot$ and the viewing angle $\phi\sim 20^\circ$,
and the latest COBE elongated bar model by \cite{bis02} with $\phi\sim 20^\circ$.

For comparisons, the right panels of Fig.~\ref{fig:opt} show the effect of removing the parallax event. The lens that caused
the parallax event is most likely in the disk. The sample without the parallax event is more representative
of the bulge structure and less influenced by the disk component, giving $\tau = [ (4.86\pm 1.94) + (1.01\pm 0.65)\times b]\times 10^{-6}$.
The gradient of $\tau$ along $b$ becomes more clearly visible in this case, as one would predict based on the GB models.

In \S\,\ref{sec:tau_blends} we use a control sample to further investigate the systematics associated
with various methods of estimating $\tau$ and discuss a possible difference between results based on
samples selected using models with and without blending.

\section{Discussion}
\label{sec:disc}

\subsection{Results from microlensing model fits}
\label{sec:fits}

Here we discuss distributions of the best fit microlensing parameters for our events. In this section event SC31-111306
is not included because it is identical with SC30-717699.

\subsubsection{Blending parameter}
\label{sec:fS}

The histograms of $f_{\rm S}$ are shown in Fig.~\ref{fig:fS} (top-left). The solid line is for 32 events in the final optical
depth sample, while the dotted line represents all 61 events at level 5. The remaining panels in this figure show the relevant
results from Monte-Carlo simulations in \S\,\ref{sec:eff}. As intended, our level 6 condition removed events
which appeared to be highly blended photometrically. In the remaining events $f_{\rm S}$ is symmetrically distributed around
the peak at $f_{\rm S}=1$.

There are several events with ``negative blending'', i.e. $f_{\rm S}>1$. It is not allowed theoretically, but also
not uncommon in fits to real data. Such unphysical best fit parameters arise due to several factors including: near degeneracy
of the model in the presence of statistical fluctuations, systematic errors in photometry (e.g. from a centroid bias or
an imperfect deblending), occasional "holes" in the background of unresolved faint stars (\citealt{par04}), or assuming an
incorrect model. In our data set the first two effects explain most cases. We allow formally $f_{\rm S}>1$ to avoid a systematic
bias in the microlensing parameters.

In Fig.~\ref{fig:Isc} (top-left) we show histograms of $I_{\rm s,0}$, the extinction corrected source magnitude.
Again, the solid line represents 32 events in the optical depth sample, while the dotted line is for all 61 events
in the level 5 sample. The remaining panels in this figure are for comparisons with Monte-Carlo simulations.
The dashed vertical line indicates the magnitude threshold $I_{\rm RC,th}$ adopted for BUL\_SC3 field.
The original goal of the blending analysis was to place a reasonable upper limit on the number of
blends and derive $\tau$ assuming $f_{\rm S}=1$ in RCG events, perhaps after applying a small blending correction
to the ingredients going into the estimator. Instead, for about half of apparently bright OGLE-II events
we found $f_{\rm S}$ and $\sigma_{f_{\rm S}+}$ that likely puts the actual source well below the magnitude threshold
$I_{\rm RC,th}$. In \S\,\ref{sec:photo_blends} this surprising observation is reconciled with simulations.

\subsubsection{Impact parameter and time-scale}
\label{sec:umin_and_te}

\begin{figure}
\begin{center}
\includegraphics[angle=-90,scale=0.7,keepaspectratio]{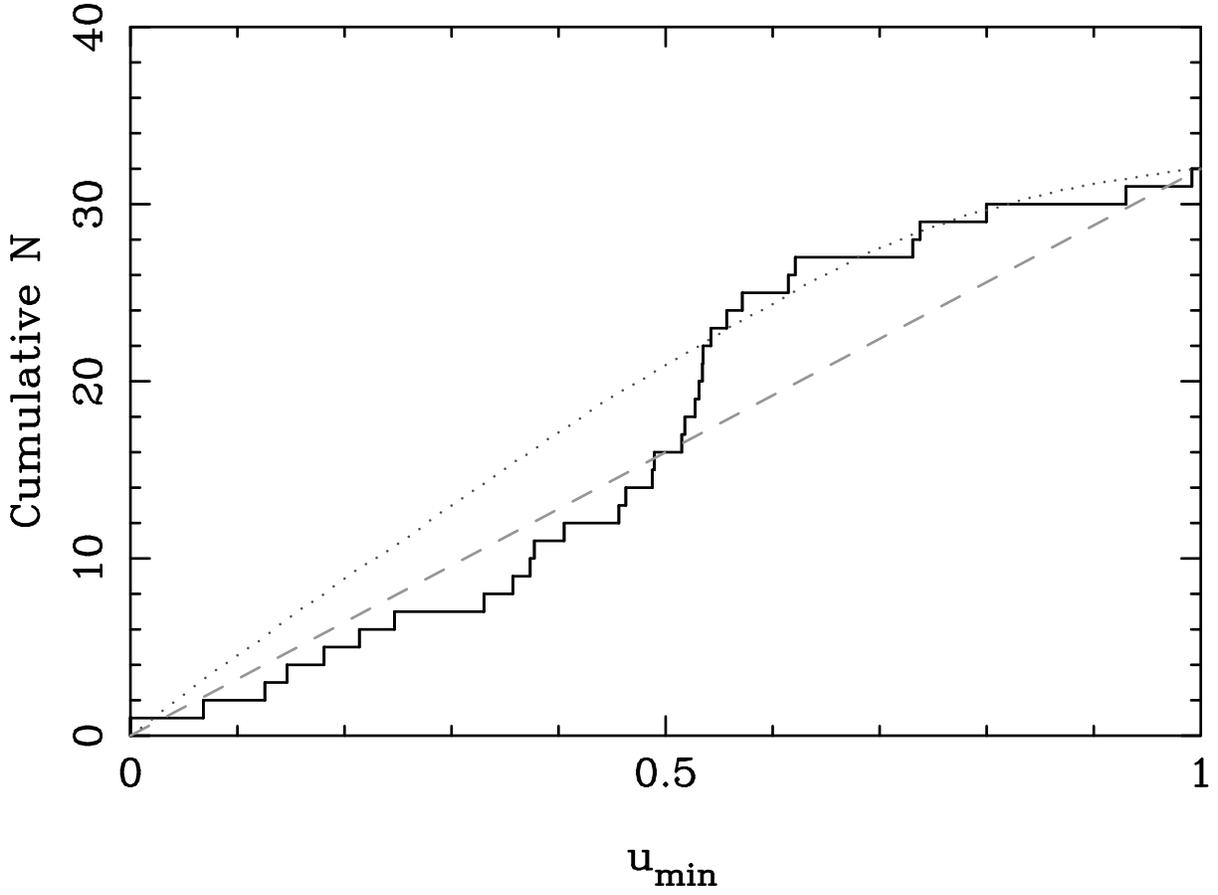}
\caption{Cumulative histogram of the best fit impact parameters $u_{\rm min}$
for 33 events in the optical depth sample. For comparison we also show a flat distribution
in $u_{\rm min}$ (dashed line) and the expected distribution from Mote-Carlo simulations (dotted line).
  \label{fig:umin} }
\end{center}
\end{figure}

In Fig.~\ref{fig:umin} we show a cumulative histogram of the best fit impact parameter for our 32 events. It shows a slight
discrepancy with respect to a uniform distribution in $u_{\rm min}$. The KS test gives a 3.4\% probability for this
departure from uniformity to occur by chance, not alarming in case of this particular type of test.
\cite{pop04} find a very similar feature. Slightly low detection efficiencies at large $u_{\rm min}$ and significant
noise in the best fit parameters while $u_{\rm min}$ is forced to be positive will tend to produce the observed pattern.

\begin{figure}
\begin{center}
\includegraphics[angle=-90,scale=0.7,keepaspectratio]{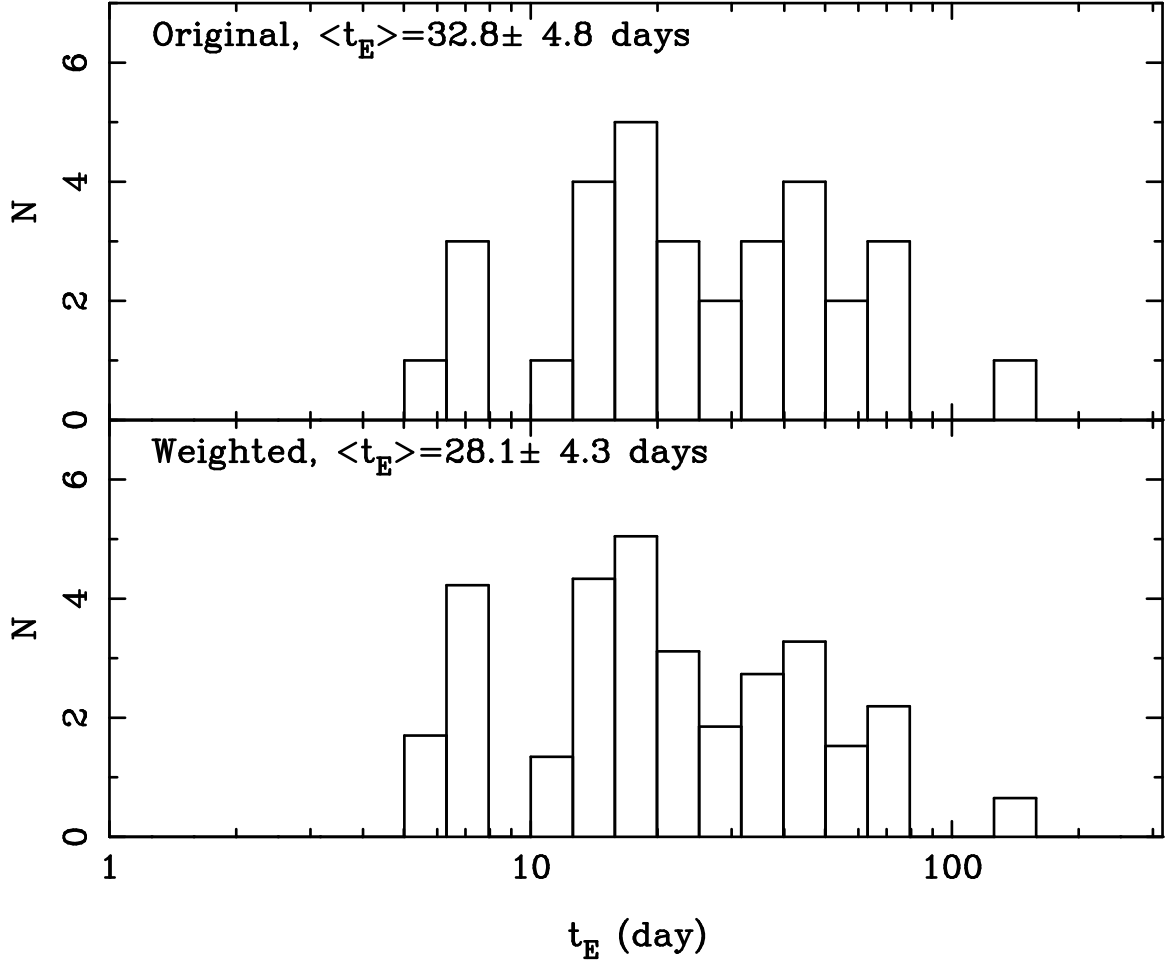}
\caption{Histogram of the best fit time-scale $t_{\rm E}$ for 32 events in the
optical depth sample. The two distributions are: uncorrected for detection efficiency (top)
and weighted by the inverse efficiency (bottom). The corrected mean time-scale $\langle t_{\rm E} \rangle =28.1\pm4.3$
d is significantly longer than 14 d predicted by the current Galactic bar models without any streaming motions.
  \label{fig:tE} }
\end{center}
\end{figure}

The corresponding histogram of $t_{\rm E}$ is shown in Fig.~\ref{fig:tE} (top).
The mean time-scale is $\langle t_{\rm E} \rangle = 32.8\pm 4.8$ d without any correction for detection efficiency.
This is close to the value of 30 d obtained by \cite{pop04}. A proper comparison with other results and 
theoretical estimates requires weighting the histogram in Fig.~\ref{fig:tE} by the inverse efficiency (bottom).
The corrected mean time-scale is $\langle t_{\rm E} \rangle =28.1\pm4.3$ d, while \cite{pop04} found $\langle t_{\rm E} \rangle = 15$ d.
\cite{woo05} have recently compared several measured time-scale distributions to predictions from a Galactic model adopted from
\cite{han03}, after it was empirically normalized by star counts. \cite{woo05} found that our corrected $t_{\rm E}$ distribution
(preprint) is in good agreement with their model, while the results in \cite{pop04} show an excess of short-duration events.
The discrepancy is likely to originate from a different treatment of blending (\S\,\ref{sec:tau_blends}).
Our corrected value is longer than the prediction from the model by \cite{bis04}. It is also below 45 d predicted by a bar model
rotating at $v_{\rm rot}=50$ km/s, but significantly longer than 14 d without any streaming motion estimated by \cite{eva02}.
Therefore, the OGLE-II data provide some support for the existence of streaming motions in the Galactic bar. 
Further analysis of the bar kinematics is beyond the scope of this paper.

\subsection{How important is blending in bright stars ?}
%==================================
\label{sec:blending}

\subsubsection{Photometric evidence for strong blends}
\label{sec:photo_blends}

As mentioned in \S\,\ref{sec:fS} the data of the kind shown in Fig.~\ref{fig:IsItotal} suggest that in many OGLE-II events
the true source is much fainter than it appears. It is somewhat unexpected that a fraction of such events should reach $\sim50$\%,
a value suggested by the number of events rejected at level 6. On closer look, however, a brief inspection of model fits for
hundreds of the OGLE-III events from the Early Warning System (EWS; \citealt{uda03}) supports this finding. As many as 20\%
of sources brighter than 16 mag are still severely blended in more densely sampled and better resolved OGLE-III images.
The results from image simulations (\S\,\ref{sec:im_eff}) help to clarify the issue.

\begin{figure*}
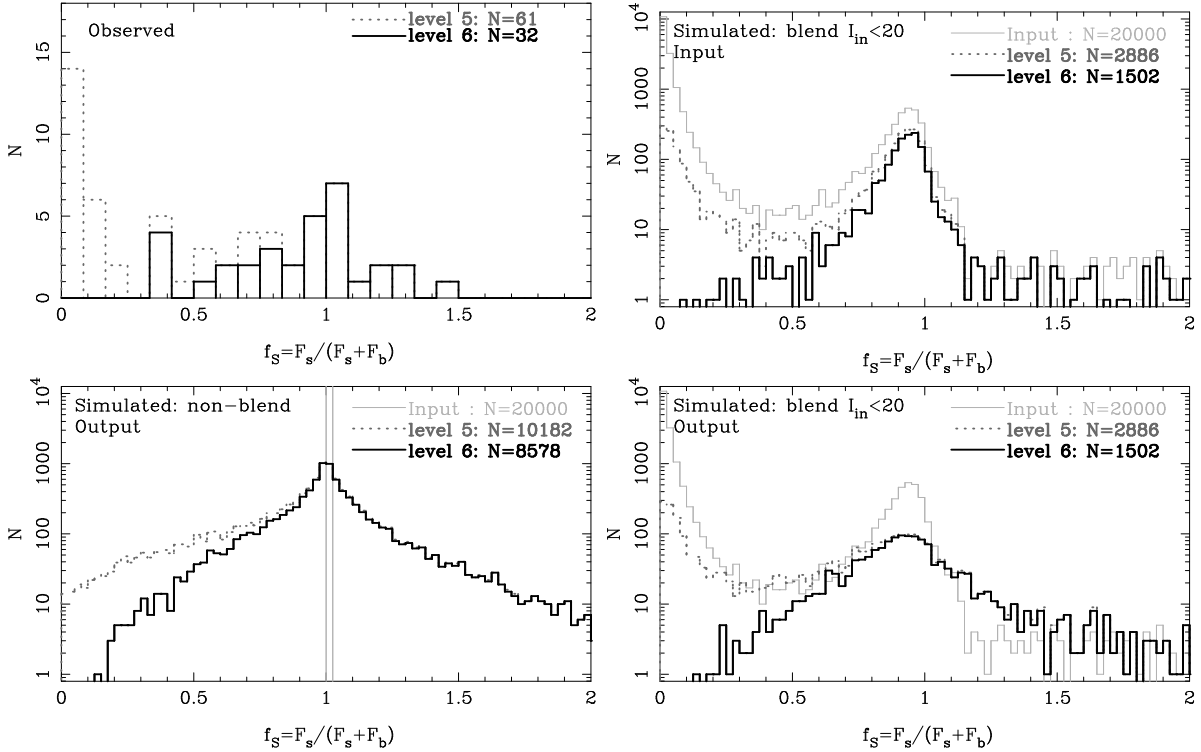

\begin{center}
\includegraphics[angle=-90,scale=0.33,keepaspectratio]{f9a.eps}
\includegraphics[angle=-90,scale=0.33,keepaspectratio]{f9b.eps}
\includegraphics[angle=-90,scale=0.33,keepaspectratio]{f9c.eps}
\includegraphics[angle=-90,scale=0.33,keepaspectratio]{f9d.eps}
 \caption{Distributions of the blending parameter $f_{\rm S}$. Solid lines are for level 6 events (optical depth sample),
and dotted lines are for level 5 events (right before rejection of blends). Light gray lines
in relevant panels represent simulated events before folding through light curve sampling
efficiency (see \S\,\ref{sec:eff} for details).
Top-left: observed samples.
Top-right: ``true'' values used to generate simulated event samples
($I_{\rm in}<20$ mag, compare Figs.~\ref{fig:IfS} and \ref{fig:histfS}).
Bottom-left:  best fit values (using a 5-parameter model with free $f_{\rm S}$)
for simulated events without blending, i.e. generated from a curve with $f_{\rm S}\equiv 1$.
The results in this panel fully support our conclusion that the observed level 5 sample
cannot be explained by a parent distribution with $f_{\rm S}=1$.
We can see that $\sim16\%$ of unblended events failed the level 6 cut.
Bottom-right: best fit values (using a 5-parameter model with free $f_{\rm S}$)
for simulated events with blending, i.e. using input $f_{\rm S}$ distribution from image simulations.
Those simulations are fully consistent with observations (Top-left panel) for both
level 5 and 6 samples.
\label{fig:fS}}
\end{center}
\end{figure*}

In the top-left panel of Fig.~\ref{fig:fS} the observed distribution of the blending parameter $f_{\rm S}$ for
level 5 events (dotted line) peaks at a low value below $0.1$. Despite a large spread in $f_{\rm S}$ introduced
by model fitting, this is inconsistent with $f_{\rm S}=1$ in all level 5 events according to the simulation results
depicted in the bottom-left panel of Fig.~\ref{fig:fS}. The opposite is true for level 6 events. The observed sample
is fully consistent with ``no blending'' scenario (solid lines).

\begin{figure*}
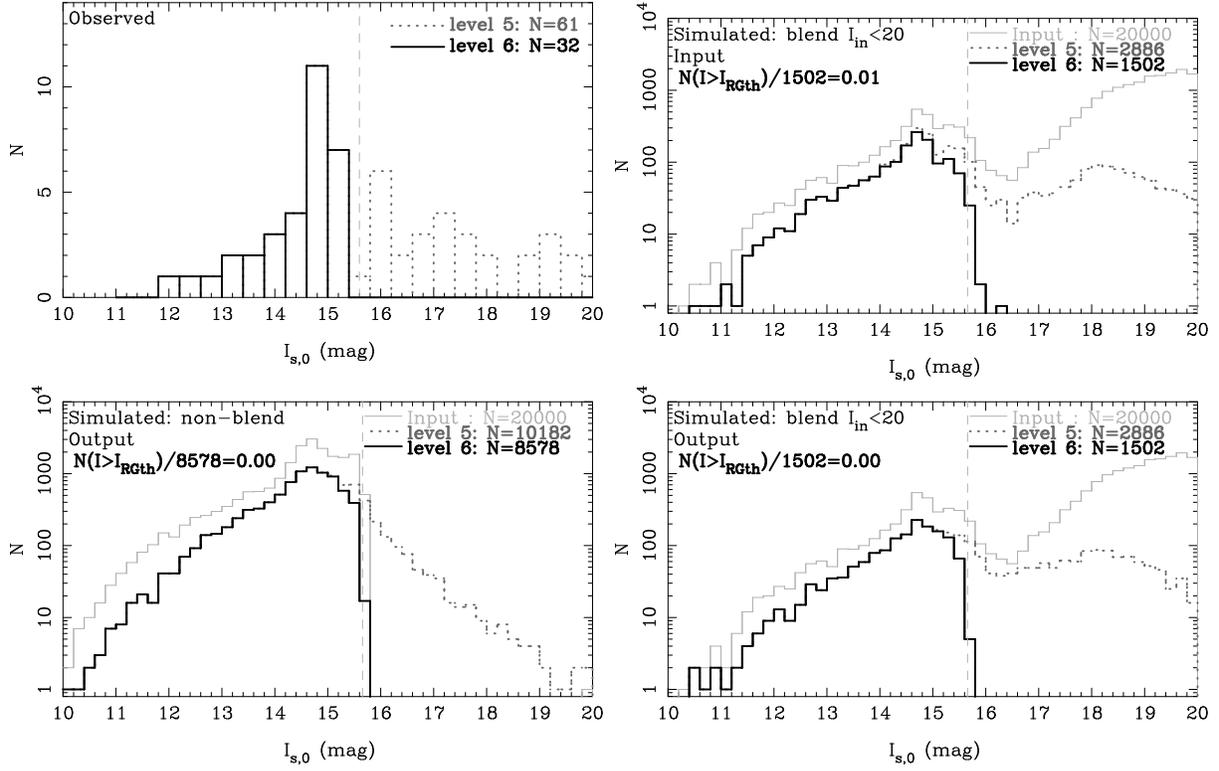

\begin{center}
\includegraphics[angle=-90,scale=0.33,keepaspectratio]{f10a.eps}
\includegraphics[angle=-90,scale=0.33,keepaspectratio]{f10b.eps}
\includegraphics[angle=-90,scale=0.33,keepaspectratio]{f10c.eps}
\includegraphics[angle=-90,scale=0.33,keepaspectratio]{f10d.eps}
\caption{Distributions of the extinction corrected source magnitude $I_{\rm s,0}$.
The assignment of samples to panels and line styles are the same as in Fig.~\ref{fig:fS}.
The dashed vertical line indicates the magnitude threshold $I_{\rm RC,th}=15.6$ in BUL\_SC3 field.
Only 1\% of the simulated level 6 events originates from source stars below the threshold (top-right).
The fraction of unblended events accidentally lost due to rejection of blends at level 6
is about 16\% (bottom-left).
  \label{fig:Isc} }
\end{center}
\end{figure*}

We also verified how efficient are level 6 criteria in rejecting most blending without removing too many unblended events.
Note that our conservative magnitude threshold employs the formal error $\sigma_{I_{\rm s+}}$ in addition to the intrinsic
source magnitude $I_{\rm s}$ (both from model fits). From Fig.~\ref{fig:Isc} we find that out of all simulated events
which make it to the optical depth sample, only 1\% has $I_{\rm s, 0} > I_{\rm RC, th}$ (top-right). At the same time only
about 16\% of unblended events are lost that way (bottom-left). The near degeneracy of the model is broken more often
with high S/N light curves delivered by DIA photometry. Some loss of efficiency was still unavoidable in cases of strong
degeneracy, because we insisted on keeping the contamination from faint blends at a low level.
Finally, the fraction of simulated events rejected for blending is very close to that in the sample of actual 
observed events, i.e. $\sim50$\% (Figs.~\ref{fig:fS} and \ref{fig:Isc}), further strenghthening the evidence for 
numerous blends.

The fraction of blended events relative to all events can be estimated using the statistics collected from the simulation.
The number of events which passed level 5 and 6 is $N_5 = 2886$ and $N_6 = 1502$, respectively (bottom-right 
panel of Fig.~\ref{fig:Isc}). Taking into account the fact that 16\% of RCG events fail the level 6 cut, the number of real
RCG events at level 5 is given by $N_{\rm cand,RCG}=1502/(1-0.16)=1788$. Then the number of blended events is given by
$N_{\rm cand,blend}=N_5 - N_{\rm cand,RCG}=1098$, which is 38\% of $N_5$. Using the observed events we have $N_{\rm cand,RCG}=38$
and $N_{\rm cand,blend}=23$, so an estimated 38\% of all events are blended events. 
This is fully consistent with our simulation.

\cite{afo03} studied blending in the EROS bulge survey data. Using a pixel level simulation they found that in the EROS bright
star sample the fraction of events due to blends is $\sim38$\%. However, the EROS analysis was based on a microlensing model
without blending, unless it produced unacceptable $\chi^2$. In \S\,\ref{sec:tau_blends} we show that this would imply even more
blended events in hypothetical EROS samples selected using a model that allows for blending. Nevertheless, both OGLE and EROS
image simulations indicate that blends are fairly common in RCG samples.

\subsubsection{Astrometric evidence for strong blends}
\label{sec:centroid_blends}

Some strong blends betray their presence by showing a detectable centroid shift of the combined light.
While in principle this information can be used to reject blended events, applying astrometric selection criteria
is complicated in practice. Another problem is that positions of some blends are aligned within the measurement errors.
Sensitivity can be improved by comparing an unbiased position from stacked difference frames with the mean centroid
of the composite. Even then, the maximum observable shift is $(1-f_{\rm S})$ times the separation of the blend
(for a baseline dominated reference frame).

\begin{figure*}
\begin{center}
\includegraphics[angle=0,scale=0.9,keepaspectratio]{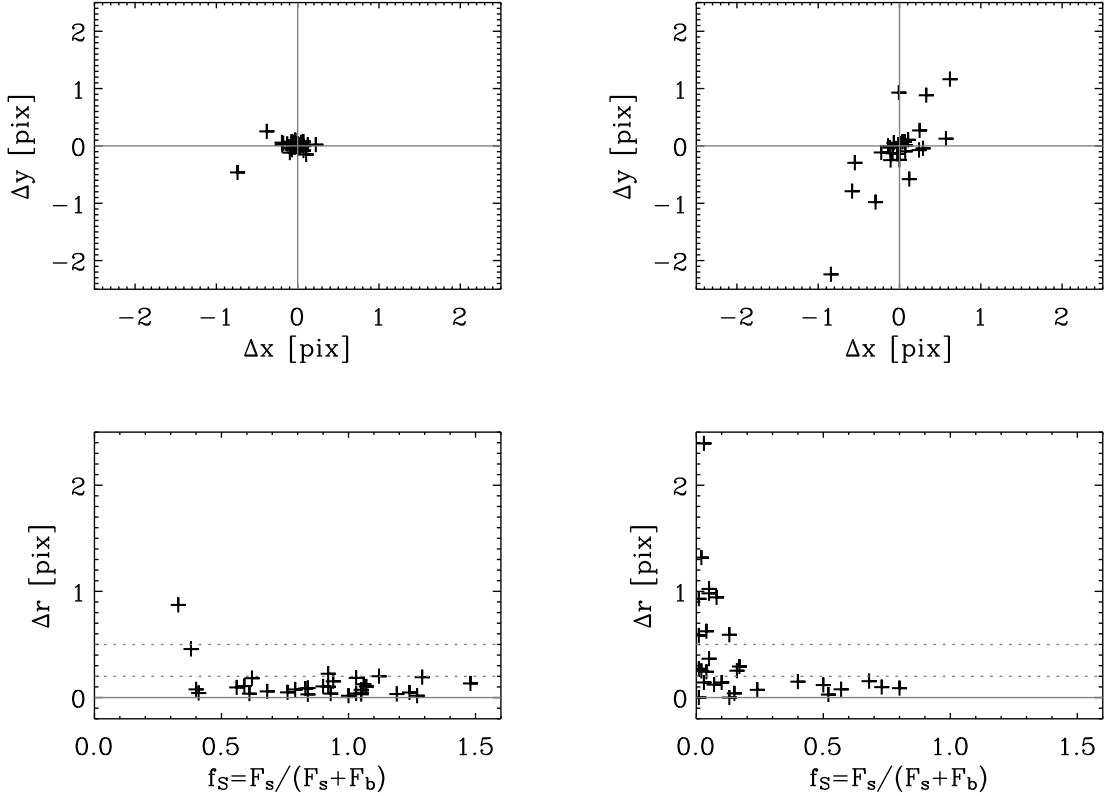}
\caption{
Centroid shifts for microlensing events selected at level 6 (left) and level 5 events rejected for blending (right).
The former sample is consistent with negligible blending, while in the full level 5 sample about one third of
all sources shows a significant centroid shift. Top panels show differences in 2D source positions between the reference
image, i.e. with low magnification, and a stack of difference images near the peak magnification. At the bottom
we plot the total shift $\Delta r$ versus the fitted fraction of the lensed light $f_{\rm S}$.
The fraction of significant centroid shifts agrees with predictions based on simulated images.
Note that large shifts are only observed in strong ``photometric blends'' with low $f_{\rm S}$.
\label{fig:astrom}
}
\end{center}
\end{figure*}

Despite these limitations we looked for astrometric shifts to check for consistency with the photometric evidence.
In Fig.~\ref{fig:astrom} we plot the differences between source positions in the reference image, typically dominated by
the baseline, and a stack of difference images taken near the peak magnification. Out of 32 level 6 events (left)
only two show a clearly detected shift above 0.2 pixels ($\sim0.08''$). That fraction increases dramatically to 28\% for
level 5 events (left and right combined) and is entirely consistent with the value $\sim30$\% predicted from a distribution of blend
separations in our simulated images with the detection efficiency as a function of $f_{\rm S}$. 
As expected, the largest shifts in Fig.~\ref{fig:astrom} are observed for strong blends with low
$f_{\rm S}$. Therefore, we have an independent confirmation that, for bright stars, the DIA photometry alone is sufficiently effective
in rejecting strong blends in OGLE-II data. Please note that the relevant selection criteria in Table~\ref{tbl:criteria}
admit blended events as long as the best fit source magnitude is $1\sigma$ above the threshold. This occurs only for a tiny fraction
of bright stars (Fig.~\ref{fig:histfS}) and has negligible influence on the optical depth.

\subsubsection{Blending, parameter bias and derived optical depths}
\label{sec:tau_blends}
                                                                                                        
In previous sections we showed that samples of microlensing events with apparently bright sources still contain many strong blends.
This holds for any survey with seeing comparable to or worse than the median ${\rm FWHM}\sim1.3''$ in OGLE-II, assuming a similar density
of stars. It may then seem counterintuitive, that our measured optical depth (adjusted for the gradient) agrees with the values recently
found by \cite{afo03} and \cite{pop04}, who used microlensing parameters based on a model without blending to select their RC samples.
To investigate this issue we reintroduced the latter method into our analysis of the OGLE-II extended RC sample. We performed microlensing model
fits with the blend flux fixed at $F_{\rm b}=0$, followed by a selection of events using the same criteria as before (Table~\ref{tbl:criteria}).
Note that the requirement on the source magnitude at level 6 is now identical to the second condition at level 0. This procedure selected 48
good quality events, all of which were confirmed visually. This control sample was not screened for blending and is 1.5 times larger than
the one used to calculate the optical depth (\S\,\ref{sec:opt}). Its sole purpose is to compare our analysis with previous work
(\citealt{afo03,pop04}), particularly to assess the level of bias in the best fit parameters and the impact on the derived optical depth.

\begin{figure}
\begin{center}
\includegraphics[angle=-90,scale=0.7,keepaspectratio]{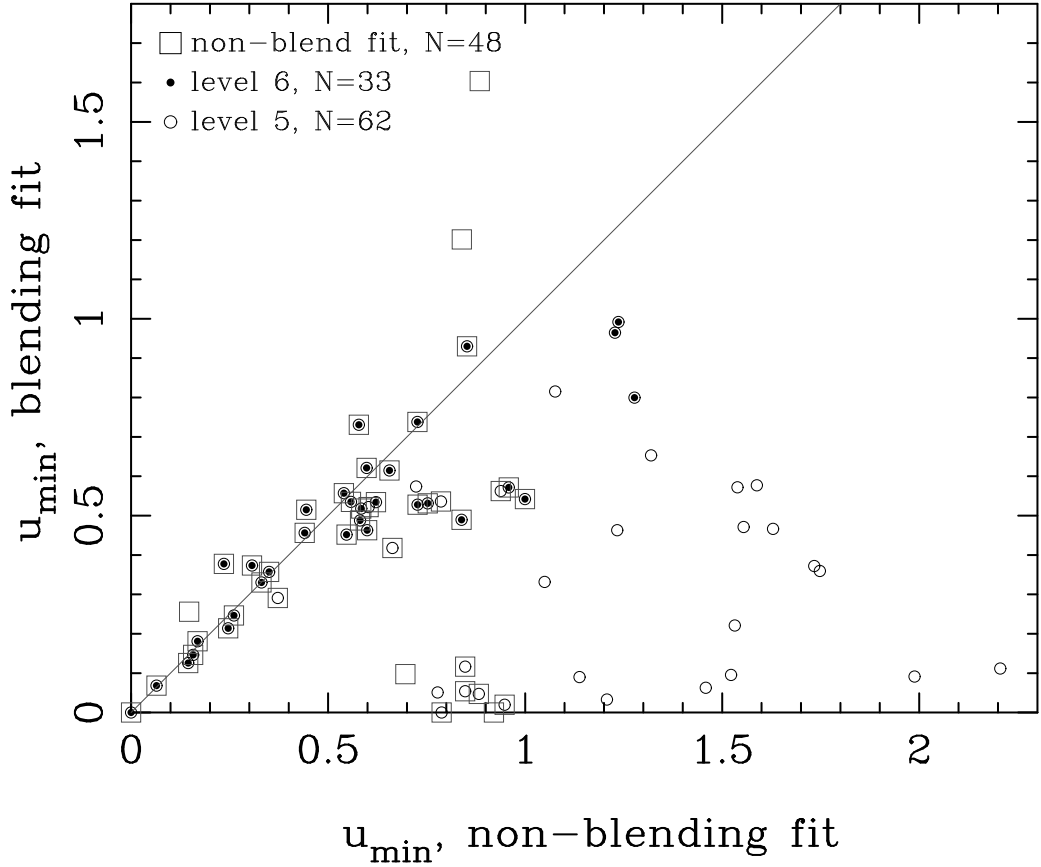}
\caption{Best fit impact parameter $u_{\rm min}$ in microlensing models with and without blending.
Compared are level 6 events (dots), level 5 events (open circles) and our control sample (open squares).
Three open squares fall outside the figure area with $u_{\rm min}$ reaching 16. Fitting a $f_{\rm S}=1$
model to significantly blended events results in overestimated $u_{\rm min}$.
  \label{fig:umin4} }
\end{center}
\end{figure}

\begin{figure}
\begin{center}
\includegraphics[angle=-90,scale=0.7,keepaspectratio]{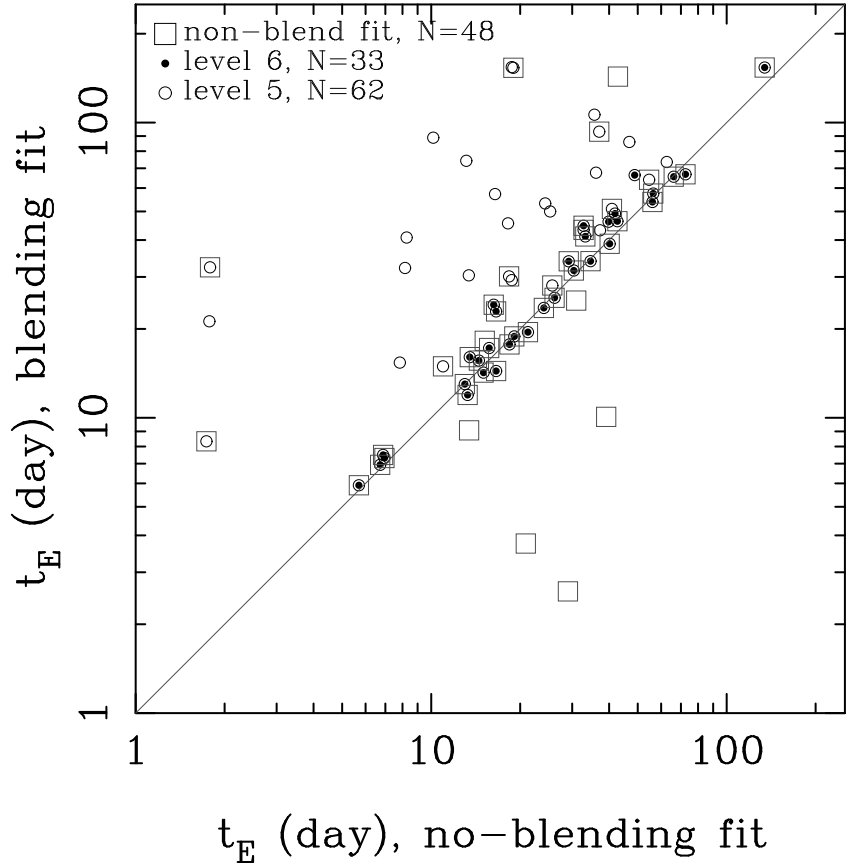}
\caption{Best fit time-scale $t_{\rm E}$ in microlensing models with and without blending.
Symbols are the same as in Fig.~\ref{fig:umin4}. One open square falls outside the figure area
with $t_{\rm E}=(3.4\pm3.4) \times 10^4$ d. Fitting a $f_{\rm S}=1$ model to significantly blended events results
in underestimated $t_{\rm E}$.
  \label{fig:tE4} }
\end{center}
\end{figure}

In Figs.~\ref{fig:umin4} and \ref{fig:tE4} we compare the results of the model fitting with and without blending.
From \S\,\ref{sec:select} we show the sample of 32 level 6 events (filled circles) and its level 5 superset (open circles).
We also plot the control sample of 48 events introduced for comparisons (open squares). There are 30 events common
between all three samples. The control sample includes 10 events rejected as blends in the optical depth analysis.
In events showing no evidence for blending there is little difference between the best fit model with fixed $f_{\rm S}=1$,
and the one with freely variable $f_{\rm S}$. For the remaining (blended) events we confirm that forcing $F_{\rm b}=0$
gives overestimated $u_{\rm min}$ and underestimated $t_{\rm E}$, a well known fact. There are 3 events in which
$t_{\rm E}$ significantly increases after fixing $f_{\rm S}=1$. Those have ``negative blends'', i.e. $f_{\rm S}>1$
(\S\,\ref{sec:fS}). Before the efficiency correction, the control sample had the mean $t_{\rm E}$ of 35.8 and 25.5 d 
corresponding to the model fits with and without blending. Event SC37-485133 has the best fit $t_{\rm E}=(3.4\pm3.4) \times 10^4$ d
when $f_{\rm S}$ is allowed to be optimized and was removed from the estimates of the mean time-scale.

%Other 3 candidates failed by the criteria $u_{\rm min}<1$.
%5 candidates with  $u_{\rm min}>1$ and 1 candidates with $t_{\rm E}>400$ 
%2 candidates without errors in parameters by
%blending fit are selected as candidates in non-blending fit.
%11 blended candidates in level 5 are also selected in  non-blending fit.

\begin{figure}
\begin{center}
\includegraphics[angle=-90,scale=0.6,keepaspectratio]{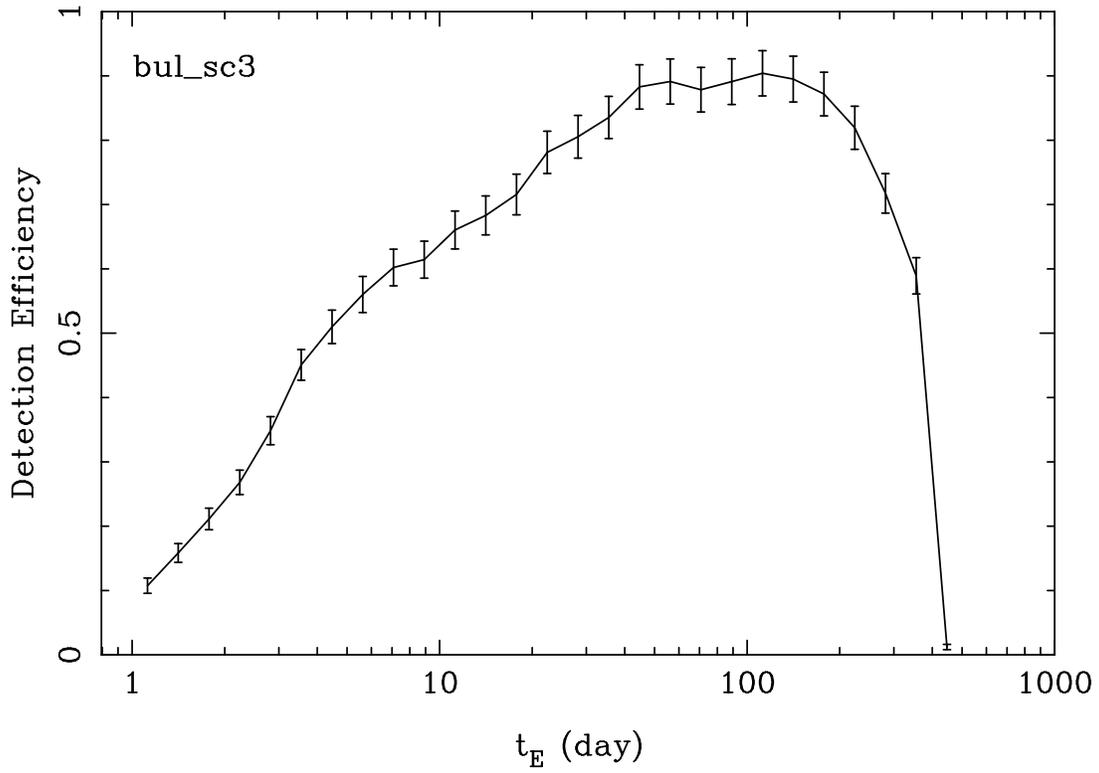}
\caption{OGLE-II microlensing detection efficiency assuming no blending in the RCG sample (BUL\_SC3).
Our data are inconsistent with that assumption. This curve is shown only for the purpose of comparing
our present analysis with previously published work.
  \label{fig:eff4} }
\end{center}
\end{figure}

The detection efficiency $\varepsilon (t_{{\rm E}})$ in the control sample was computed following the methods in \S\,\ref{sec:eff}
except that simulated events were assumed to have $f_{\rm S}\equiv1$, as in \cite{afo03} and \cite{pop04}. The result for BUL\_SC3
field is shown in Fig.~\ref{fig:eff4}. Assuming no blending, the efficiency is 30\% higher on average compared to
$\varepsilon (t_{{\rm E}})$ in a blended scenario (Fig. \ref{fig:eff}), 
largely because fixing the blending parameter amounts to less degeneracy.

The optical depth estimate with the control sample is consequently 
$\tau = (1.97 \pm 0.35) \times 10^{-6}$, where the error is from \cite{han95} formula. 
Although $1\sigma$ lower than our main result, this value explicitly demonstrates a near cancellation of the two major competing biases
discussed by \cite{afo03} and \cite{pop04}. The efficiency corrected mean time-scale in the control sample
is $\langle t_{\rm E} \rangle =21.9\pm3.2$ d, shorter by 20\% than $\langle t_{\rm E} \rangle =28.1\pm4.3$ d estimated in
\S\,\ref{sec:umin_and_te}. Event selection using model fits with $f_S\equiv 1$ produced a sample of roughly 50\% more events
with $t_{\rm E}$ underestimated by $\sim20$\% and $\varepsilon (t_{{\rm E}})$ higher by $\sim30$\%. Although the latter
assumption is incorrect, it leads to a final value of the optical depth that is statistically indistinguishable
from the one we found using a proper model. Note that an estimated 2/5 out of 50\% more events stated above are real RCG events,
knowing that 16\% of real RCG events fail the level 6 cut (see \S \ref{sec:photo_blends}). So, the fraction of blended events
in the control sample is $\sim 20$\% (30/150), consistent with 19\% from Fig. \ref{fig:umin4} and similar to 17\%
estimated by \cite{pop04} in MACHO RCG sample.

The difference between optical depth estimates based on two different treatments of blending in our data is at an inconclusive $1\sigma$ level.
Generally, it is possible that one of these methods of estimating $\tau$ is more biased than the other. Such bias might have contributed
to somewhat higher $\tau$ values from analyses based on blended microlensing fits (\citealt{alc00b,sume03} and \S\,\ref{sec:opt} in this paper),
compared to values based on fits without blending (\citealt{pop04,afo03} and our control sample). The issue warrants further investigation,
however finding a reliable answer will likely require very detailed simulations that are beyond the scope of this paper.

It should be emphasized how risky is the argument that unblended fit can be used when introducing it into the model
does not significantly lower the $\chi^2$. The more blending in the event, the more degenerate is the model fit.
With a typical S/N stronger blends are actually harder to find using $\chi^2$ improvements. We have evidence for heavy
degeneracy in 12 of our level 4 candidates, and mild degeneracy in many more cases. Despite having little effect
on the optical depth estimate, neglecting blends biases time-scale distributions and event rates. 
The degree of cancellation likely depends on the specifics of each dataset, e.g. typical seeing, and eventually 
will no longer be ``perfect'' as statistical errors continue to shrink.

\subsubsection{Effects of weak parallax and binary events on optical depth}
\label{sec:weakParaBin}

In low S/N events, a 5-parameter model fit with blending is substantially more degenerate than
a corresponding 4-parameter fit with $f_{\rm S}\equiv 1$, and can be sensitive to small changes in the data.
Such low level light curve deviations may be caused by weak exotic events, most commonly by parallax
and binary lens events. The problem with this type of deviations is that they are hard to include in
simulated events, so they are typically ignored in the detection efficiency calculations. It is an open question
whether the presence of those weakly exotic events in samples selected using a 5-parameter model
introduces a bias in the optical depth. The issue has not been sufficiently studied. A correction for such
hypothetical bias could in principle be obtained from extensive simulations that account for the Galactic
Disk geometry, but such calculations are not available today.

There is some evidence that weakly exotic events are not significantly biasing our measurement.
A thorough search for parallax events (including weak signals) has been performed by \cite{smi02} using the
first 3 years of the OGLE-II data. Out of 512 candidate microlensing events, a single convincing event
(herein SC33-553617) and 5 more marginal parallax events were found (after excluding two events that are actually
binaries). The fraction is about 1 \%.
Parallax events are most likely due to disk lenses whose fraction is rather uncertain and estimated at
10--20 \% from disk models. Moreover, the expected time-scale is 50 days or more, so only 10--20 \% of all events
are relevant here. Therefore, we expect that only a few per cent of the events have parallax effects.
Knowing that 1 event has already been found in the sample, we expect negligible contribution from parallax
events beyond what is already included.

In case of binary events, the binary fraction from published work is 3--8 \%, and we correct the
measurement using our own estimated 6 \% fraction. We do not expect more significant binary lens
events. Any bias in the optical depth due to a combination of subtle light curve changes and fitting
5-parameter models would have to be accounted for. However, to our best knowledge, such bias has not
been reported.

\section{Summary and conclusions}
%==================================
\label{sec:summary}

We found 81 microlensing events in a sample of about 1 million bright source stars in the extended RCG region.
Light curves for 62 of those events are well represented by a single lens model with relatively small uncertainty in the
best fit parameters. In the latter sample we investigated blending and concluded that about 38\% of microlensing events
with the apparent RCG sources are actually due to fainter stars below the magnitude threshold. This large value for the
fraction of blended events is fully supported by our extensive Monte-Carlo simulations of source confusion in the OGLE-II
reference images. 

The microlensing optical depth was estimated using 32 high S/N events whose source stars 
are still in the extended RCG region. We measured 
%$\tau = 2.37_{-0.43}^{+0.53} \times 10^{-6}$ at $(l,b)=(1.^\circ16, -2.^\circ75)$. 
$\tau=2.55_{-0.46}^{+0.57}\times 10^{-6}$ at $(l,b)=(1.^\circ16, -2.^\circ75)$.
Considering our estimate of the gradient along the Galactic latitude $b$,
%$\tau = [ (4.17\pm 2.20) + (0.73\pm 0.78)\times b]\times 10^{-6} $, 
$\tau = [ (4.48\pm 2.37) + (0.78\pm 0.84)\times b]\times 10^{-6}$,
this value is consistent with recent 
measurements based on RCG source stars by \cite{pop04} and \cite{afo03}.

Our goal was to ensure a high quality measurement of $\tau$ using RCG stars in the Galactic Bulge. The result presented here
is based on a rigorous treatment of source confusion and a relatively large number of well sampled high S/N events,
with the purpose of minimizing uncertainties in $t_{\rm E}$ and in the number of events. It may therefore seem surprising that our
result is not out of line with the values obtained by \cite{pop04} and \cite{afo03}, who selected RCG samples using an unblended
model. On the other hand, these authors have also argued that the two dominant biases nearly cancel each other.
The contribution due to an underestimated
number of monitored sources has the opposite sign to that from underestimated time-scales in blending free model fits.
We tested the hypothesis of near cancellation of the bias by reanalyzing our microlensing sample. Indeed, under the no blending
assumption we found nearly the same value of $\tau = (1.97 \pm 0.35) \times 10^{-6}$ with roughly 50\% more events, 20\% lower $t_{\rm E}$
and 30\% higher detection efficiency. This does not change the fact that such procedure shifts the distribution of time-scales
to lower values by significant amount and makes it hard to study the kinematics of the source/lens populations.

The efficiency weighted mean time-scale in our sample of 32 events is $\langle t_{\rm E} \rangle =28.1\pm4.3$ d,
which is significantly longer than 15 d estimated by \cite{pop04}.
This is not as long as 45 d predicted by models with the streaming motion of the bar, but significantly longer than 14 d
without any streaming (\citealt{eva02}). Therefore, our result implies some streaming in the bar (first measured by
\citealt{sumEW03}). Further analysis of the time-scale distribution is planned including the information in the 
OGLE-II proper motion catalog (\citealt{sume04}).

Our value of $\tau$ agrees with predictions from \cite{han95} model with the bulge mass $M_{\rm bulge}=1.62\times 10^{10} M_\odot$ and
the viewing angle $\phi\sim 20^\circ$. It is also consistent with microlensing optical depths derived by \cite{eva02} using Galactic bulge
models of \cite{dwe95} and \cite{bin97}. 

It has been noted that $\tau$ measurements based on RCG stars (\citealt{afo03,pop04}) fall systematically below the estimates using 
all sources down to the detection limit, including recent DIA results (\citealt{alc00b,sume03}). The difference persists, although
at a lower level, when the dubious correction for disk stars is not applied. We are not aware of a convincing detailed explanation
of this gap, although source confusion and model degeneracy biasing $t_{\rm E}$ and $u_{\rm min}$ are prime suspects.
Alternatively, the difference may be related to the treatment of blending. The significance of any potential biases due to near
degeneracy of the 5-parameter microlensing curve should be established. The systematics of event selection using model fits
with free $f_{\rm S}$ (\citealt{alc00b,sume03}, and this work) and with $f_{\rm S}\equiv 1$ (\citealt{afo03,pop04}) may be the culprit.
Another area of concern for wider use of the blended microlensing curve is a possible perturbing effect on the optical depth
due to the presence of unrecognized weakly exotic events discussed in \S\,\ref{sec:tau_blends} and \S\,\ref{sec:weakParaBin}.

The event selection criteria in this analysis strongly discriminate against a high level of degeneracy in the fit (\S\,\ref{sec:level345}).
It is our goal to estimate $\tau$ using all events of any magnitude and investigate the source of the above discrepancy. A good understanding
of the involved statistics will allow tapping into a much larger set of useful events. As of 2005, in the GB fields the OGLE-III survey
detects about 600 events per year. According to \cite{han95} one needs roughly 700--800 events to distinguish a barred Galactic bulge from
its axisymmetric alternative, and with $\sim 1800$ events we can hope for a 3$\sigma$ measurement of the relative contributions from
the bulge and disk components. Future statistical modeling of the microlensing survey data needs to reflect source confusion effects,
still evident among bright stars.

%The candidate list is available in electronic format via anonymous ftp from 
%{\it ftp://ftp.astrouw.edu.pl/ogle/ogle2/extinction/} or
%{\it ftp://bulge.princeton.edu/ogle/ogle2/extinction/}.

\acknowledgments
We acknowledge A. Gould, O. Gerhard and M. Smith for helpful comments.
TS acknowledges the financial support from the JSPS. PW was supported by the Oppenheimer Fellowship at LANL.
This work was partly supported with the following grants to BP: NSF grant AST-0204908,
and NASA grant NAG5-12212. The OGLE project is partly supported by the Polish KBN grant 2P03D02124 to AU.

\clearpage

%\normalsize

%----------------------------- FIG. 1 -------------------------------------
%--------------------------------------------------------------------------
%----------------------------- FIG. 2 -------------------------------------
%--------------------------------------------------------------------------

%----------------------------- FIG. 3 -------------------------------------
%--------------------------------------------------------------------------

%----------------------------- FIG. 4 -------------------------------------
\begin{figure*}
\begin{center}
\includegraphics[angle=0,scale=0.8,keepaspectratio]{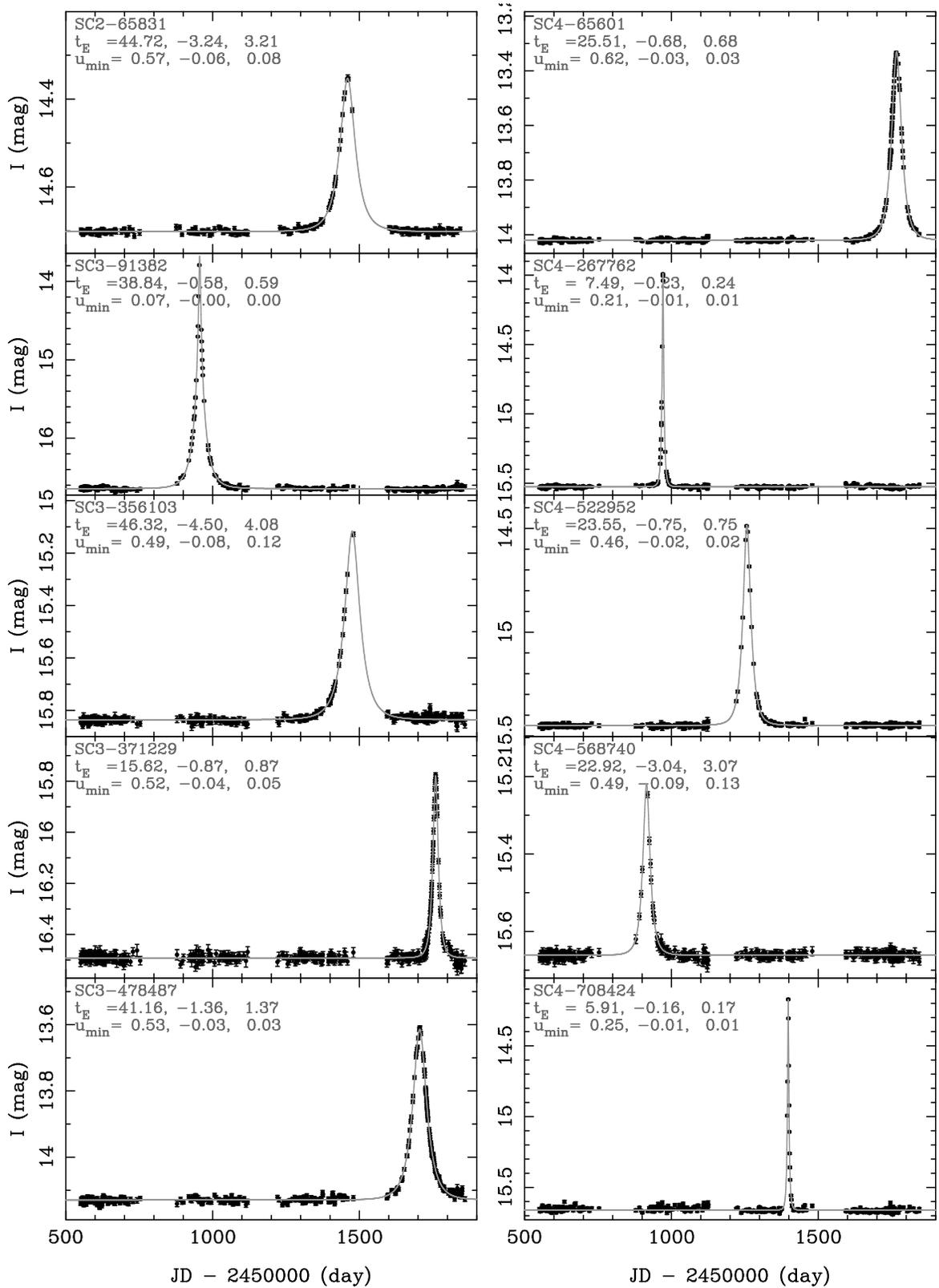}
\caption{Light curves of 33 level 6 events. This is our optical depth sample.
Note that SC31-111306 and SC30-717699 are the same event. The field/star ID,
as well as the best fit parameters $t_{\rm E}$ and $u_{\rm min}$ with their
68\% lower and upper confidence limits are also shown. For SC33-553617, the 
best fit parallax model is shown as dotted line.
\label{fig:lc} }
\end{center}
\end{figure*}
%--------------------------------------------------------------------------
\clearpage
\begin{figure*}
\begin{center}
\includegraphics[angle=0,scale=0.8,keepaspectratio]{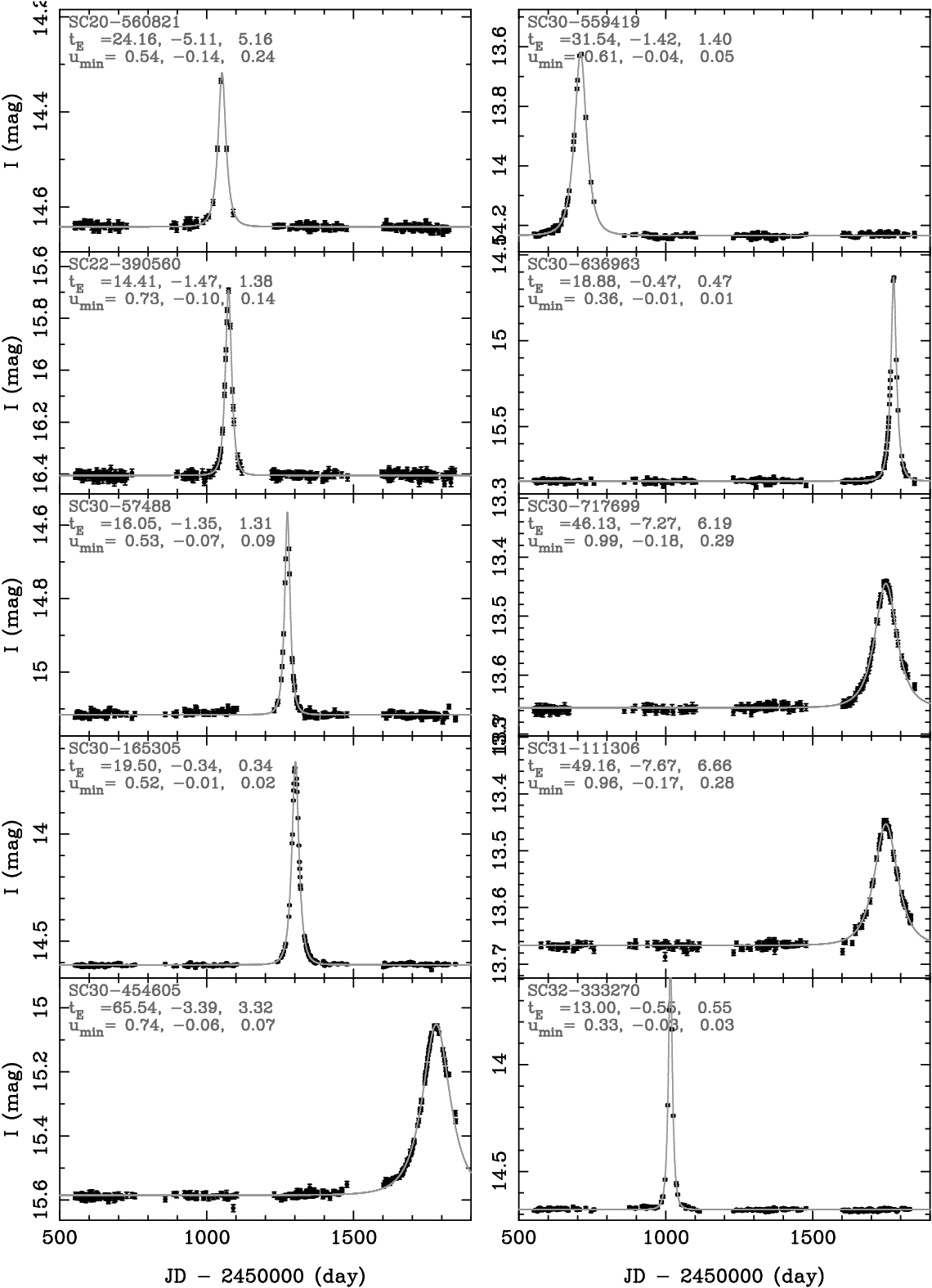}
\contcaption{continued.}
\end{center}
\end{figure*}
%--------------------------------------------------------------------------
\clearpage
\begin{figure*}
\begin{center}
\includegraphics[angle=0,scale=0.8,keepaspectratio]{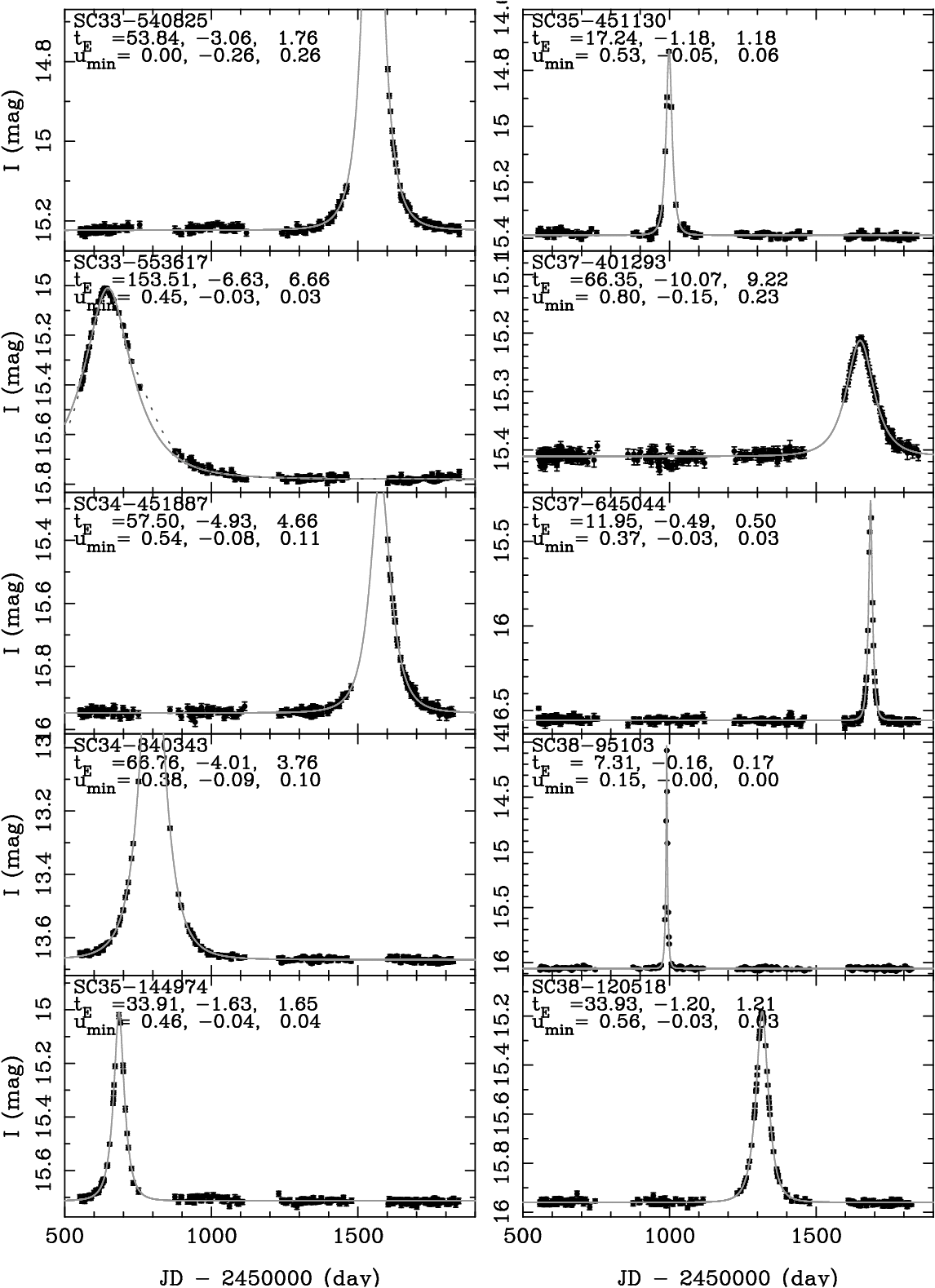}
\contcaption{continued.}
\end{center}
\end{figure*}
%--------------------------------------------------------------------------
\clearpage
\begin{figure}
\begin{center}
\includegraphics[angle=0,scale=0.8,keepaspectratio]{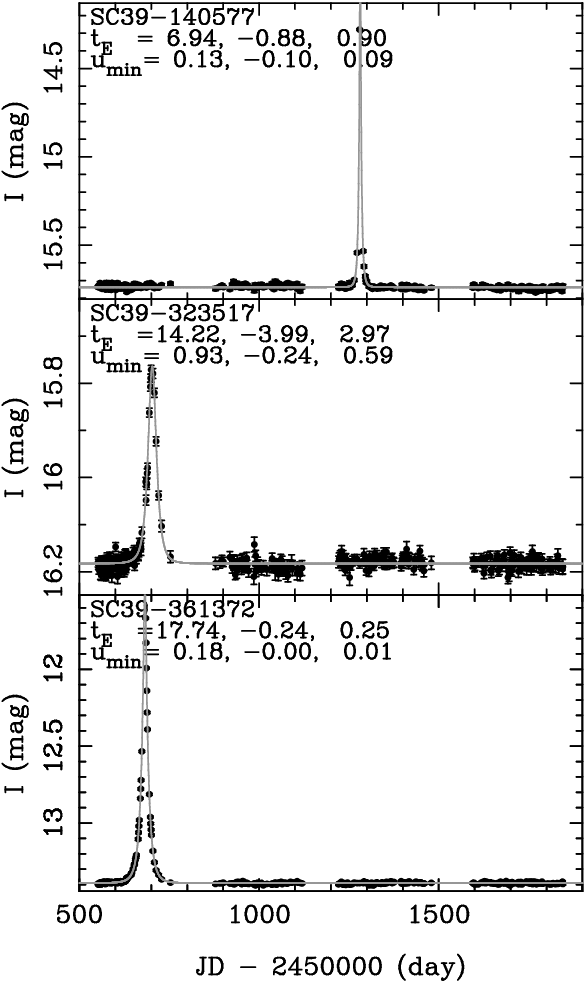}
\contcaption{continued.}
\end{center}
\end{figure}
%--------------------------------------------------------------------------

%--------------------------------FIG. 5-------------------------------------
%--------------------------------------------------------------------------

%----------------------------- FIG. 6 -------------------------------------
%--------------------------------------------------------------------------
%----------------------------- FIG. 7 -------------------------------------
%--------------------------------------------------------------------------

%----------------------------- FIG. 8 -------------------------------------
%--------------------------------------------------------------------------

%----------------------------- FIG. 9 -------------------------------------
%--------------------------------------------------------------------------

%----------------------------- FIG. 10 -------------------------------------
%--------------------------------------------------------------------------

%----------------------------- FIG. 11 -------------------------------------
%--------------------------------------------------------------------------
%----------------------------- FIG. 12 -------------------------------------
%--------------------------------------------------------------------------

%----------------------------- FIG. 13 -------------------------------------
%--------------------------------------------------------------------------

%----------------------------- FIG. 14 -------------------------------------
%--------------------------------------------------------------------------

%----------------------------- FIG. 15 -------------------------------------
%--------------------------------------------------------------------------

%----------------------------- FIG. 16 -------------------------------------
%--------------------------------------------------------------------------

%----------------------------- FIG. 17 -------------------------------------
%--------------------------------------------------------------------------

%----------------------------- FIG. 18 -------------------------------------
%--------------------------------------------------------------------------

%------------------------Table 1.---------------------------------
\begin{deluxetable}{clccccc}
\tablecaption{OGLE-II GB fields contributing to the present optical depth analysis. Also given are
   Galactic coordinates of the field center ($l$, $b$), number of source stars ($N_{\rm s}$),
   number of microlensing events ($N_{\rm lens}$), optical depth ($\tau$) and its error ($\sigma_{\tau}$). \label{tbl:fld}}
\tablewidth{0pt}
\tablehead{
    \colhead{Field} & \colhead{$l$ ($^\circ$)} & \colhead{$b$ ($^\circ$)} & \colhead{$N_{\rm s}$} &
    \colhead{$N_{\rm lens}$}  & \colhead{$\tau (10^{-6})$} & \colhead{$\sigma_{\tau} (10^{-6})$}
}
\startdata
 1 &  1.08 & -3.62 &   35844 &  0 & 0.000000 & 0.000000 \\
 2 &  2.23 & -3.46 &   38278 &  1 & 2.483758 & 2.483758 \\
 3 &  0.11 & -1.93 &   80976 &  4 & 4.286616 & 2.226285 \\
 4 &  0.43 & -2.01 &   78454 &  5 & 3.145996 & 1.494732 \\
20 &  1.68 & -2.47 &   58891 &  1 & 1.239517 & 1.239517 \\
21 &  1.80 & -2.66 &   52930 &  0 & 0.000000 & 0.000000 \\
22 & -0.26 & -2.95 &   50732 &  1 & 0.854807 & 0.854807 \\
23 & -0.50 & -3.36 &   42205 &  0 & 0.000000 & 0.000000 \\
30 &  1.94 & -2.84 &   47974 &  6 & 9.543987 & 4.183375 \\
31 &  2.23 & -2.94 &   45825 &  1 & 2.261020 & 2.261020 \\
32 &  2.34 & -3.14 &   40476 &  1 & 0.937948 & 0.937948 \\
33 &  2.35 & -3.66 &   34461 &  2 & 10.415521 & 7.933656 \\
34 &  1.35 & -2.40 &   61515 &  2 & 4.087111 & 2.890040 \\
35 &  3.05 & -3.00 &   41146 &  2 & 3.211394 & 2.350345 \\
36 &  3.16 & -3.20 &   37922 &  0 & 0.000000 & 0.000000 \\
37 &  0.00 & -1.74 &   85289 &  2 & 2.211746 & 1.772319 \\
38 &  0.97 & -3.42 &   40148 &  2 & 2.876526 & 2.246087 \\
39 &  0.53 & -2.21 &   72842 &  3 & 1.621968 & 0.967699 \\
45 &  0.98 & -3.94 &   32823 &  0 & 0.000000 & 0.000000 \\
46 &  1.09 & -4.14 &   29626 &  0 & 0.000000 & 0.000000 \\
\enddata
\end{deluxetable}

\clearpage
%------------------------Table 2.---------------------------------
\begin{deluxetable}{llr}
\tablecaption{Selection criteria for microlensing events. The last column gives the number of candidate events passed to the next level.
 \label{tbl:criteria}}
\tablewidth{0pt}
\tablehead{
    \colhead{Level} &  \colhead{Criteria} & \colhead{$N_{\rm cand}$}
}
\startdata
0  &  $I_0 < 9 \times (V-I)_0 + I_{\rm RC,0}-5.5 $                &           \\
   &  $I_0 < I_{\rm RC, th}$                                      &           \\
   &  $N_{\rm data} \ge 70$                                       & 1,084,267 \\
1  &  $0<N_{\rm peak} <4$                                         &           \\
   &  $\sigma_{\rm max} \ge 6$                                    &           \\
   &  $\sum_{\rm peak,max} \sigma_i \ge 20$                       &           \\
   &  $\chi_{\rm out}^2/d.o.f < 2.2$ for $\sigma_{\rm max} <10$   & 821       \\
%
%2  &  $520 \le t_{0}\le 1870$                                     &           \\
2  &  $530 \le t_{0}\le 1860$                                     &           \\
   &  $t_{\rm E}\le 400$ d                                        &           \\
   &  $u_{\rm min} <1$                                            &           \\
   &  $\chi^2_{\rm ml} \le 1.0$ for $\sigma_{\rm max}< 7.5$       &           \\
   &  $\chi^2_{\rm ml} \le 1.5$ for $7.5\le \sigma_{\rm max}< 10$ &           \\
   &  $\chi^2_{\rm ml} \le 1.8$ for $10 \le \sigma_{\rm max}< 30$ & 135       \\
3  &  reject spurious events                                      & 81        \\
4  &  error estimates converged for all parameters                & 69        \\
5  &  $\chi^2_{\rm ml} \le 3.5$ and no cross ref with binary events & 62        \\
6  &  $I_{\rm s,0} + \sigma_{I_{\rm s+}} < I_{\rm RC, th}$        & 33        \\
\enddata
\end{deluxetable}

%------------------------Table 3.---------------------------------
\begin{deluxetable}{lll}
\tablecaption{ Binary lens events cross-referenced with other samples. \label{tbl:binary}}
\tablewidth{0pt}
\tablehead{
    \colhead{ID$_{\rm OGLE}$} & \colhead{ID$_{\rm Ref}$} & \colhead{Reference}
}
\startdata
SC20-395103 & 401.48408.649  or 98-BLG-14 & \cite{alc00a} \\
SC30-352272 & 108.19333.1878 or 108-E     & \cite{alc00a} \\
SC21-45456  & 113.18674.756  or 97-BLG-1  & \cite{alc00a} \\
SC20-69068  & sc20-1793                   & \cite{jar02}  \\
\enddata
\end{deluxetable}

\clearpage
%------------------------Table 4.---------------------------------
\begin{deluxetable}{lccrrrlccc}
\tabletypesize{\scriptsize}
\tablecaption{Microlensing events used in the optical depth measurement (after the final level 6 cut).
Only 32 events are unique due to a duplicate event in the overlap region between two fields: SC31-111306 and SC30-717699.\label{tbl:candlist6}
}
\tablewidth{0pt}
\tablehead{
\colhead{Field-ID} & 
\colhead{R.A.} & 
\colhead{Dec.} & 
\colhead{$\sigma_{\rm max}$} & 
\colhead{$t_0\ \ \ \ \ \ $}  & 
\colhead{$t_{\rm E}\ \ \ \ $} & 
\colhead{$\ \ u_{\rm min}$} & 
\colhead{$f_{\rm S}$} & 
\colhead{$I_{\rm s,0}$} & 
\colhead{$\frac{\chi^2}{dof}$} 
%\colhead{$\frac{\chi^2_{\rm peak}}{dof}$} 
}
\startdata
  SC2-65831 & 18:04:06.85 & $-$29:01:17.2 &   57 &  1459.9$\pm0.4$ &  44.7$\pm3.1$ & 0.572 $\pm0.066$ &   0.40$\pm0.07$ & 15.0$\pm0.2$ & 0.56\\
  SC3-91382 & 17:53:09.33 & $-$30:01:12.1 &  135 &   956.0$\pm0.0$ &  38.8$\pm0.6$ & 0.068 $\pm0.002$ &   1.05$\pm0.02$ & 14.9$\pm0.0$ & 0.95\\
 SC3-356103 & 17:53:33.93 & $-$29:46:15.6 &   54 &  1475.8$\pm1.0$ &  46.3$\pm4.2$ & 0.488 $\pm0.095$ &   0.76$\pm0.22$ & 14.7$\pm0.3$ & 0.90\\
 SC3-371229 & 17:53:21.13 & $-$29:40:37.4 &   49 &  1759.6$\pm0.1$ &  15.6$\pm0.9$ & 0.518 $\pm0.046$ &   0.83$\pm0.11$ & 15.0$\pm0.1$ & 0.76\\
 SC3-478487 & 17:53:47.38 & $-$30:05:26.2 &   56 &  1705.5$\pm0.1$ &  41.2$\pm1.4$ & 0.531 $\pm0.029$ &   0.56$\pm0.05$ & 13.4$\pm0.1$ & 1.64\\
  SC4-65601 & 17:54:09.41 & $-$29:53:36.9 &  111 &  1767.1$\pm0.0$ &  25.5$\pm0.7$ & 0.621 $\pm0.026$ &   1.06$\pm0.07$ & 12.8$\pm0.1$ & 0.60\\
 SC4-267762 & 17:54:21.79 & $-$29:53:24.0 &  147 &   971.1$\pm0.0$ &   7.5$\pm0.2$ & 0.214 $\pm0.011$ &   0.84$\pm0.05$ & 14.6$\pm0.1$ & 1.03\\
 SC4-522952 & 17:54:38.64 & $-$29:33:12.8 &  121 &  1257.1$\pm0.1$ &  23.6$\pm0.7$ & 0.456 $\pm0.024$ &   1.05$\pm0.08$ & 14.1$\pm0.1$ & 0.55\\
 SC4-568740 & 17:54:49.35 & $-$29:20:25.0 &   39 &   915.2$\pm0.2$ &  22.9$\pm3.0$ & 0.490 $\pm0.108$ &   0.41$\pm0.13$ & 15.2$\pm0.4$ & 0.53\\
 SC4-708424 & 17:55:00.07 & $-$29:35:03.7 &  126 &  1397.8$\pm0.0$ &   5.9$\pm0.2$ & 0.247 $\pm0.010$ &   0.93$\pm0.05$ & 14.5$\pm0.1$ & 2.35\\
SC20-560821 & 17:59:27.19 & $-$28:32:31.5 &   42 &  1052.2$\pm0.3$ &  24.2$\pm5.2$ & 0.542 $\pm0.183$ &   0.33$\pm0.17$ & 14.9$\pm0.6$ & 1.01\\
SC22-390560 & 17:56:56.81 & $-$31:11:51.0 &   48 &  1074.3$\pm0.1$ &  14.4$\pm1.4$ & 0.731 $\pm0.122$ &   1.48$\pm0.43$ & 14.9$\pm0.3$ & 0.70\\
 SC30-57488 & 18:01:02.52 & $-$29:00:11.6 &   64 &  1274.8$\pm0.1$ &  16.0$\pm1.3$ & 0.528 $\pm0.082$ &   0.61$\pm0.14$ & 14.8$\pm0.2$ & 0.98\\
SC30-165305 & 18:01:07.74 & $-$28:31:41.7 &  155 &  1302.0$\pm0.0$ &  19.5$\pm0.3$ & 0.515 $\pm0.015$ &   1.24$\pm0.05$ & 13.5$\pm0.0$ & 1.61\\
SC30-454605 & 18:01:30.89 & $-$28:59:26.7 &   61 &  1781.4$\pm0.2$ &  65.5$\pm3.1$ & 0.738 $\pm0.056$ &   1.03$\pm0.14$ & 14.6$\pm0.2$ & 1.35\\
SC30-559419 & 18:01:33.95 & $-$28:28:02.3 &   81 &   709.9$\pm0.2$ &  31.5$\pm1.4$ & 0.615 $\pm0.047$ &   0.90$\pm0.11$ & 13.4$\pm0.1$ & 1.62\\
SC30-636963 & 18:01:44.80 & $-$28:58:03.5 &  113 &  1775.7$\pm0.0$ &  18.9$\pm0.5$ & 0.357 $\pm0.015$ &   1.03$\pm0.06$ & 14.9$\pm0.1$ & 0.84\\
SC30-717699 & 18:01:52.15 & $-$28:32:36.5 &   38 &  1749.3$\pm0.3$ &  46.1$\pm6.6$ & 0.992 $\pm0.217$ &   0.62$\pm0.28$ & 13.3$\pm0.5$ & 0.76\\
SC31-111306 & 18:01:52.14 & $-$28:32:36.5 &   36 &  1749.6$\pm0.3$ &  49.2$\pm7.1$ & 0.965 $\pm0.211$ &   0.59$\pm0.26$ & 13.4$\pm0.5$ & 0.96\\
SC32-333270 & 18:03:21.74 & $-$28:28:50.4 &  108 &  1015.8$\pm0.0$ &  13.0$\pm0.5$ & 0.330 $\pm0.031$ &   1.00$\pm0.10$ & 13.9$\pm0.1$ & 1.07\\
SC33-540825 & 18:05:45.79 & $-$28:30:52.4 &   45 &  1552.0$\pm0.5$ &  53.8$\pm1.7$ & 0.000 $\pm0.273$ &   1.12$\pm0.10$ & 14.2$\pm0.1$ & 0.85\\
SC33-553617 & 18:05:46.71 & $-$28:25:32.1 &   74 &   647.1$\pm0.3$ &  153.5$\pm6.7$ & 0.452 $\pm0.028$ &   0.75$\pm0.07$ & 15.3$\pm0.1$ & 2.76\\
SC33-553617\tablenotemark{a} & -- & -- &   -- &   668.3$\pm1.1$ &  140.9$\pm16.4$ & 0.405 $\pm0.101$ &   0.84$\pm0.26$ & 15.2$\pm0.4$ & 1.21\\
SC34-451887 & 17:58:14.20 & $-$28:48:22.3 &   45 &  1576.3$\pm1.5$ &  57.5$\pm4.9$ & 0.535 $\pm0.098$ &   0.94$\pm0.25$ & 14.7$\pm0.3$ & 0.78\\
SC34-840343 & 17:58:37.12 & $-$29:06:29.9 &   62 &   798.9$\pm0.3$ &  66.8$\pm3.9$ & 0.377 $\pm0.096$ &   1.29$\pm0.24$ & 12.3$\pm0.2$ & 0.95\\
SC35-144974 & 18:04:09.65 & $-$27:44:34.9 &   54 &   684.9$\pm0.1$ &  33.9$\pm1.6$ & 0.463 $\pm0.038$ &   0.68$\pm0.08$ & 15.2$\pm0.1$ & 0.85\\
SC35-451130 & 18:04:33.63 & $-$28:07:32.2 &   68 &   998.6$\pm0.1$ &  17.2$\pm1.1$ & 0.534 $\pm0.057$ &   0.79$\pm0.13$ & 14.8$\pm0.2$ & 1.61\\
SC37-401293 & 17:52:32.30 & $-$29:58:46.8 &   22 &  1650.3$\pm0.5$ &  66.3$\pm9.5$ & 0.800 $\pm0.179$ &   0.38$\pm0.16$ & 14.6$\pm0.5$ & 0.82\\
SC37-645044 & 17:52:58.69 & $-$29:34:22.2 &   78 &  1685.7$\pm0.0$ &  12.0$\pm0.5$ & 0.373 $\pm0.029$ &   1.27$\pm0.13$ & 14.5$\pm0.1$ & 0.91\\
 SC38-95103 & 18:01:09.74 & $-$29:56:18.9 &  197 &   990.5$\pm0.0$ &   7.3$\pm0.2$ & 0.146 $\pm0.005$ &   0.92$\pm0.03$ & 15.3$\pm0.0$ & 1.00\\
SC38-120518 & 18:01:10.23 & $-$29:48:55.2 &   88 &  1316.1$\pm0.1$ &  33.9$\pm1.2$ & 0.557 $\pm0.031$ &   1.05$\pm0.09$ & 15.0$\pm0.1$ & 0.79\\
SC39-140577 & 17:55:17.08 & $-$29:37:41.0 &  150 &  1281.4$\pm0.1$ &   6.9$\pm0.9$ & 0.126 $\pm0.083$ &   0.92$\pm0.29$ & 14.7$\pm0.4$ & 1.11\\
SC39-323517 & 17:55:36.42 & $-$29:42:14.0 &   29 &   702.7$\pm0.2$ &  14.2$\pm3.3$ & 0.930 $\pm0.333$ &   1.19$\pm0.85$ & 14.8$\pm0.9$ & 0.89\\
SC39-361372 & 17:55:28.67 & $-$29:33:41.7 &  154 &   682.3$\pm0.0$ &  17.7$\pm0.2$ & 0.181 $\pm0.005$ &   1.07$\pm0.03$ & 12.1$\pm0.0$ & 0.65\\
\enddata
\tablecomments{The symmetric 68\% confidence intervals shown here are only for the purpose of quick reference and avoiding a large amount
of unnecessary details. The analysis is based on proper asymmetric confidence intervals. Complete information is available electronically at ApJ web.\\}
\tablenotetext{a}{Parallax model fit with $\psi= 3.17\pm0.12$ radian and  $\tilde{r}_{\rm E}=6.16\pm0.39$ AU.}
\end{deluxetable}

\clearpage

%------------------------Table 5.---------------------------------
\begin{deluxetable}{lccrrrlccc}
\tabletypesize{\scriptsize}
%\rotate
\tablecaption{Microlensing events with evidence for faint blended sources. Those 29 events were rejected from the optical depth sample at level 6.
 \label{tbl:candlist5}}
\tablewidth{0pt}
\tablehead{
\colhead{Field-ID} &
\colhead{R.A.} &
\colhead{Dec.} &
\colhead{$\sigma_{\rm max}$} &
\colhead{$t_0\ \ \ \ \ \ $}  &
\colhead{$t_{\rm E}\ \ \ \ $} &
\colhead{$\ \ u_{\rm min}$} &
\colhead{$f_{\rm S}$} &
\colhead{$I_{\rm s,0}$} &
\colhead{$\frac{\chi^2}{dof}$}
}
\startdata
 SC3-147406 & 17:53:09.61 & $-$29:46:39.3 &   14 &  1409.2$\pm0.4$ &  57.3$\pm10.6$ & 0.221 $\pm0.058$ &   0.04$\pm0.01$ & 17.4$\pm0.4$ & 1.20\\
 SC3-576464 & 17:53:41.56 & $-$29:39:06.0 &   40 &   656.5$\pm0.2$ &  43.3$\pm2.2$ & 0.418 $\pm0.035$ &   0.50$\pm0.06$ & 15.7$\pm0.1$ & 0.76\\
 SC3-577610 & 17:53:36.20 & $-$29:38:08.2 &   23 &  1779.7$\pm0.1$ &  74.2$\pm13.3$ & 0.051 $\pm0.012$ &   0.04$\pm0.01$ & 19.2$\pm0.2$ & 0.69\\
 SC3-601945 & 17:53:36.63 & $-$29:31:21.2 &   17 &   924.4$\pm0.5$ &  43.2$\pm9.3$ & 0.574 $\pm0.193$ &   0.68$\pm0.36$ & 16.1$\pm0.6$ & 1.05\\
 SC4-134300 & 17:54:13.04 & $-$29:35:14.1 &   21 &  1041.0$\pm0.2$ &  153.3$\pm50.0$ & 0.047 $\pm0.018$ &   0.02$\pm0.01$ & 19.3$\pm0.4$ & 0.82\\
 SC4-321697 & 17:54:25.13 & $-$29:37:49.4 &   36 &  1615.6$\pm0.2$ &  28.1$\pm2.5$ & 0.522 $\pm0.075$ &   0.80$\pm0.17$ & 15.8$\pm0.2$ & 0.77\\
 SC4-489287 & 17:54:42.92 & $-$29:44:12.1 &   15 &  1751.5$\pm0.1$ &  15.4$\pm3.1$ & 0.463 $\pm0.130$ &   0.16$\pm0.07$ & 17.2$\pm0.5$ & 0.80\\
 SC4-500461 & 17:54:41.53 & $-$29:40:10.1 &   41 &   745.8$\pm0.3$ &  30.1$\pm3.5$ & 0.000 $\pm0.232$ &   0.17$\pm0.03$ & 16.4$\pm0.4$ & 0.53\\
 SC4-624085 & 17:55:00.19 & $-$29:59:53.6 &   14 &   659.3$\pm0.3$ &  29.3$\pm4.1$ & 0.653 $\pm0.149$ &   0.24$\pm0.09$ & 15.8$\pm0.5$ & 1.73\\
 SC4-719953 & 17:54:56.68 & $-$29:31:47.6 &   37 &  1725.8$\pm0.1$ &  32.3$\pm8.2$ & 0.020 $\pm0.006$ &   0.01$\pm0.00$ & 19.7$\pm0.3$ & 0.66\\
SC20-525747 & 17:59:28.45 & $-$28:42:53.3 &   14 &  1746.3$\pm0.2$ &   5.3$\pm3.9$ & 0.063 $\pm0.140$ &   0.02$\pm0.02$ & 19.0$\pm1.1$ & 0.66\\
SC21-766993 & 18:00:39.38 & $-$28:55:14.6 &   28 &  1119.6$\pm1.5$ &  93.2$\pm16.3$ & 0.116 $\pm0.047$ &   0.07$\pm0.02$ & 17.6$\pm0.4$ & 0.68\\
SC22-380074 & 17:56:58.83 & $-$31:14:05.6 &   16 &   587.2$\pm0.4$ &  40.8$\pm16.7$ & 0.090 $\pm0.053$ &   0.03$\pm0.02$ & 19.3$\pm0.7$ & 0.74\\
SC22-414328 & 17:56:59.20 & $-$31:03:22.3 &   31 &  1832.8$\pm0.6$ &  51.0$\pm5.1$ & 0.536 $\pm0.092$ &   0.52$\pm0.14$ & 16.2$\pm0.3$ & 0.62\\
SC23-524386 & 17:57:56.24 & $-$30:51:58.9 &   82 &  1434.8$\pm0.1$ &  64.0$\pm1.6$ & 0.291 $\pm0.011$ &   0.73$\pm0.03$ & 15.9$\pm0.1$ & 1.02\\
SC30-671185 & 18:01:47.58 & $-$28:49:04.9 &   22 &   557.6$\pm0.6$ &  50.0$\pm6.6$ & 0.332 $\pm0.068$ &   0.15$\pm0.04$ & 17.4$\pm0.3$ & 1.05\\
SC32-208566 & 18:03:22.47 & $-$29:03:43.2 &   11 &   690.0$\pm2.0$ &  67.6$\pm20.3$ & 0.577 $\pm0.296$ &   0.13$\pm0.10$ & 17.3$\pm1.2$ & 0.74\\
   SC33-505 & 18:05:03.99 & $-$29:18:08.6 &   14 &  1778.5$\pm0.4$ &  30.4$\pm6.5$ & 0.466 $\pm0.153$ &   0.08$\pm0.04$ & 17.1$\pm0.6$ & 0.84\\
SC33-290665 & 18:05:29.67 & $-$28:51:03.3 &   22 &  1845.3$\pm4.3$ &  73.5$\pm18.9$ & 0.815 $\pm0.374$ &   0.57$\pm0.50$ & 15.8$\pm1.4$ & 0.73\\
SC34-606996 & 17:58:34.18 & $-$29:06:29.4 &    8 &   974.9$\pm0.9$ &  154.1$\pm71.0$ & 0.091 $\pm0.053$ &   0.01$\pm0.00$ & 19.5$\pm0.6$ & 0.72\\
SC37-556534 & 17:53:02.92 & $-$30:03:07.8 &   30 &  1786.3$\pm0.0$ &   8.3$\pm2.3$ & 0.054 $\pm0.031$ &   0.05$\pm0.02$ & 17.8$\pm0.4$ & 0.83\\
  SC39-1073 & 17:55:10.06 & $-$30:10:29.5 &   24 &  1807.5$\pm0.2$ &  14.9$\pm3.3$ & 0.563 $\pm0.205$ &   0.40$\pm0.23$ & 16.3$\pm0.7$ & 0.87\\
 SC39-28566 & 17:55:10.31 & $-$30:05:11.7 &   14 &  1374.0$\pm0.2$ &  21.2$\pm11.4$ & 0.033 $\pm0.023$ &   0.01$\pm0.01$ & 20.0$\pm0.7$ & 0.75\\
 SC39-54533 & 17:55:19.53 & $-$29:57:57.0 &    9 &  1223.9$\pm2.2$ &  45.6$\pm15.5$ & 0.372 $\pm0.210$ &   0.05$\pm0.04$ & 17.7$\pm1.3$ & 0.62\\
SC39-269524 & 17:55:23.99 & $-$29:55:33.1 &   11 &   906.7$\pm0.6$ &  32.2$\pm9.1$ & 0.096 $\pm0.099$ &   0.03$\pm0.01$ & 16.9$\pm0.5$ & 1.79\\
SC39-322789 & 17:55:26.87 & $-$29:43:59.7 &   12 &   968.0$\pm0.8$ &  53.2$\pm12.1$ & 0.471 $\pm0.162$ &   0.10$\pm0.05$ & 16.9$\pm0.6$ & 0.75\\
SC39-468687 & 17:55:45.16 & $-$29:56:38.6 &   10 &  1210.8$\pm3.7$ &  86.0$\pm22.2$ & 0.572 $\pm0.256$ &   0.13$\pm0.10$ & 16.8$\pm1.0$ & 0.82\\
SC39-576039 & 17:55:43.97 & $-$29:24:28.7 &    7 &   620.1$\pm1.1$ &  106.4$\pm40.5$ & 0.360 $\pm0.188$ &   0.05$\pm0.03$ & 18.1$\pm0.8$ & 1.04\\
SC39-753576 & 17:56:01.00 & $-$29:25:11.9 &    6 &  1241.2$\pm0.9$ &  88.9$\pm35.8$ & 0.112 $\pm0.056$ &   0.01$\pm0.00$ & 19.0$\pm0.6$ & 0.79\\
\enddata
\tablecomments{The symmetric 68\% confidence intervals shown here are only for the purpose of quick reference and avoiding a large amount
of unnecessary details. The analysis is based on proper asymmetric confidence intervals. Complete information is available electronically at ApJ web.}
\end{deluxetable}

%------------------------------------------------------------------

\end{document}